\documentclass[a4paper,11pt]{article}
\pdfoutput=1 

\usepackage{jheppub} 

\usepackage[T1]{fontenc} 
\usepackage[english]{babel}
\usepackage{graphicx}			    	
\usepackage{slashed}
\usepackage{caption}
\usepackage{amsmath}
\usepackage{amsthm}
\usepackage{amsfonts}
\usepackage{mathtools}
\usepackage{subcaption}
\usepackage{tikz} 
\usepackage{tikz-feynman}
\usepackage{tikz-3dplot}
\tikzfeynmanset{compat=1.1.0}
\usepackage{placeins}
\usepackage{scalerel}
\usepackage{verbatim}
\usepackage{underscore}
\usetikzlibrary{shapes.misc}
\tikzset{cross/.style={cross out, draw=black, minimum size=2*(#1-\pgflinewidth), inner sep=0pt, outer sep=0pt},
	cross/.default={5pt}}

\DeclareMathOperator{\Det}{Det}
\DeclareMathOperator{\Tr}{Tr}

\newcommand*\Laplace{\mathop{}\!\mathbin\bigtriangleup}
\theoremstyle{plain}
\theoremstyle{plain}

\theoremstyle{definition}

\makeatletter
\def\@fpheader{\relax}
\makeatother

\title{UV asymptotics of $n$-point correlators of twist-$2$ operators in SU($N$) Yang-Mills theory}

\author[a]{Marco Bochicchio}
\author[b,a]{Mauro Papinutto}
\author[b,a]{Francesco Scardino}

\affiliation[a]{Physics Department, INFN Roma1, \\
Piazzale A. Moro 2, Roma, I-00185, Italy}
\affiliation[b]{Physics Department, Sapienza University,\\Piazzale A. Moro 2, Roma, I-00185, Italy}

\emailAdd{marco.bochicchio@roma1.infn.it}
\emailAdd{mauro.papinutto@roma1.infn.it}
\emailAdd{francesco.scardino@roma1.infn.it}

\abstract{The generating functional $\mathcal{W}^E[J_{\mathcal O}]$ of Euclidean correlators of twist-$2$ operators in SU($N$) Yang-Mills theory admits the 't Hooft large-$N$ expansion:
$\mathcal{W}^E[J_{\mathcal O}]=\mathcal{W}^E_{sphere}[J_{\mathcal O}]+\mathcal{W}^E_{torus}[J_{\mathcal O}]+ \cdots$.
Nonperturbatively, $\mathcal{W}^E_{sphere}[J_{\mathcal O}]$ is a sum of tree diagrams involving glueball propagators and vertices, while $\mathcal{W}^E_{torus}[J_{\mathcal O}]$ is a sum of glueball one-loop diagrams. Moreover, it has been predicted that $\mathcal{W}^E_{torus }[J_{\mathcal O}]$ should admit the structure of the logarithm of a functional determinant summing glueball one-loop diagrams.
We work out in a closed form the ultraviolet (UV) asymptotics of $\mathcal{W}^E_{sphere}[J_{\mathcal O},\lambda] \sim \mathcal{W}^E_{asym \, sphere}[J_{\mathcal O},\lambda]$ and $\mathcal{W}^E_{torus}[J_{\mathcal O},\lambda]  \sim \mathcal{W}^E_{asym \, torus}[J_{\mathcal O},\lambda]$ in the coordinate representation as all the coordinates of the correlators are uniformly rescaled by a factor $\lambda \rightarrow 0$.
The calculation is performed in two steps.
First, extending our previous work, we compute -- directly from its path-integral definition as a Gaussian integral -- the generating functional of the conformal correlators $\mathcal{W}_{conf}[J_{\mathcal O}]=\mathcal{W}_{conf \, sphere}[J_{\mathcal O}]+\mathcal{W}_{conf \, torus}[J_{\mathcal O}]$ to the lowest perturbative order of all the twist-$2$ operators with maximal spin along the $p_+$ direction, both in Minkowskian and -- by analytical continuation -- Euclidean space-time.
Thus, we provide a purely perturbative explanation as to why $\mathcal{W}_{conf}[J_{\mathcal O}]$ has the structure of the logarithm of a functional determinant. Second, by means of a careful choice of the renormalization scheme that reduces the mixing of the above operators to the multiplicatively renormalizable case to all orders of perturbation theory, we lift the generating functional of the Euclidean conformal correlators $\mathcal{W}^E_{conf}[J_{\mathcal O}]$ to the generating functional of the renormalization-group improved correlators $\mathcal{W}^E_{asym}[J_{\mathcal O},\lambda]=\mathcal{W}^E_{asym \, sphere}[J_{\mathcal O},\lambda]+\mathcal{W}^E_{asym \, torus}[J_{\mathcal O},\lambda]$ that inherits the very same structure of the logarithm of a functional determinant. 
Remarkably, we verify the above prediction that $\mathcal{W}^E_{asym \, torus}[J_{\mathcal O},\lambda]$ -- being asymptotic in the UV to $\mathcal{W}^E_{torus}[J_{\mathcal O}, \lambda]$ -- admits the structure of the logarithm of a functional determinant as well. Hence, the computation above sets strong UV asymptotic constraints on the nonperturbative solution of large-$N$ YM theory and it may be a pivotal guide for the search of such a solution.}

\begin{document} 

\definecolor{c969696}{RGB}{150,150,150}
\maketitle	
\flushbottom

  \section{Introduction, physics motivations and conclusions} \label{s1}

The generating functional $\mathcal{W}^E[J_{\mathcal O}]=\mathcal{W}^E[J_{\mathbb{O}},J_{\tilde{\mathbb{O}}},J_{\mathbb{S}},J_{\bar{\mathbb{S}}}]$ of Euclidean connected correlators of single-trace gauge-invariant twist-$2$ operators in SU($N$) Yang-Mills (YM) theory admits the 't Hooft large-$N$ expansion \cite{tHooft:1973alw} in powers of $\frac{1}{N}$:
\begin{equation}
\mathcal{W}^E[J_{\mathcal O}]=\mathcal{W}^E_{sphere}[J_{\mathcal O}]+\mathcal{W}^E_{torus}[J_{\mathcal O}]+ \cdots
\end{equation}
Perturbatively, in terms of the 't Hooft gauge coupling $g^2=N g^2_{YM}$ \cite{tHooft:1973alw}, $\mathcal{W}^E_{sphere}[J_{\mathcal O}]$ contains a sum of Feynman diagrams that -- in the 't Hooft double-line representation -- have the topology of a punctured sphere, while $\mathcal{W}^E_{torus}[J_{\mathcal O}]$ includes diagrams that have the topology of a punctured torus, the punctures arising in both cases from the insertion of the sources $J_{\mathcal O}$ dual to the twist-$2$ operators $\mathcal{O}$. \par
Alternatively, $\mathcal{W}^E_{sphere}[J_{\mathcal O}]$ may be defined as a sum of planar diagrams \cite{tHooft:1973alw} and $\mathcal{W}^E_{torus}[J_{\mathcal O}]$ as a sum of leading-order (LO) nonplanar contributions.\par
Nonperturbatively, in terms of the renormalization-group (RG) invariant scale $\Lambda_{YM}$, $\mathcal{W}^E_{sphere}[J_{\mathcal O}]$ is a sum of tree diagrams involving glueball propagators and vertices \cite{Witten:1979kh}, while $\mathcal{W}^E_{torus}[J_{\mathcal O}]$ is a sum of glueball one-loop diagrams, both arising from the effective description of the -- yet to come --  nonperturbative solution of large-$N$ YM theory as a theory of an infinite number of weakly coupled glueballs with coupling of order $\frac{1}{N}$ \cite{tHooft:1973alw, Migdal:1977nu,Witten:1979kh}.\par
Moreover, on the basis of the existence of a nonperturbative glueball effective action, it has been predicted \cite{Bochicchio:2016toi} that $\mathcal{W}^E_{torus }[J_{\mathcal O}]$ should admit the structure of the logarithm of a functional determinant summing glueball one-loop diagrams.\par
Presently, nothing is known quantitatively on the structure of $\mathcal{W}^E_{sphere}[J_{\mathcal O}]$ and $\mathcal{W}^E_{torus}[J_{\mathcal O}]$ nonperturbatively.
Hence, one of the aims of the present paper is to put strong quantitative constraints on the aferomentioned nonperturbative structure by working out in a closed form the ultraviolet (UV) asymptotics of $\mathcal{W}^E_{sphere}[J_{\mathcal O},\lambda] \sim \mathcal{W}^E_{asym \, sphere}[J_{\mathcal O},\lambda]$ and $\mathcal{W}^E_{torus}[J_{\mathcal O},\lambda]  \sim \mathcal{W}^E_{asym \, torus}[J_{\mathcal O},\lambda]$ in the coordinate representation as all the coordinates of the correlators are uniformly rescaled by a factor $\lambda \rightarrow 0$. This is achieved in two steps:\par
(I) The efficient computation in a closed form of the generating functional of conformal correlators $\mathcal{W}^E_{conf}[J_{\mathcal O}] = \mathcal{W}^E_{conf \,sphere}[J_{\mathcal O}]+ \mathcal{W}^E_{conf \, torus}[J_{\mathcal O}]$ of twist-$2$ operators to the lowest order of perturbation theory.\par
(II) The RG-improvement from $\mathcal{W}^E_{conf}[J_{\mathcal O}] = \mathcal{W}^E_{conf \,sphere}[J_{\mathcal O}]+ \mathcal{W}^E_{conf \, torus}[J_{\mathcal O}]$ to $\mathcal{W}^E_{asym}[J_{\mathcal O},\lambda] = \mathcal{W}^E_{asym \,sphere}[J_{\mathcal O},\lambda]+ \mathcal{W}^E_{asym \, torus}[J_{\mathcal O},\lambda]$, which is obtained by standard RG-methods in combination with a new technique \cite{MB1} that reduces the operator mixing of twist-$2$ operators to the multiplicatively renormalizable case to all orders of perturbation theory.\par
We describe the steps (I) and (II) in more detail as follows.\par
Recently, we have computed \cite{BPS1} to the lowest perturbative order in YM theory $n$-point conformal correlators in the coordinate $G^{(n)}_{conf}(x_1,\ldots,x_n)$ and momentum $G^{(n)}_{conf}(p_1,\ldots,p_n)$ representation of the gauge-invariant twist-$2$ operators with maximal spin along the $p_+$ direction, both in Minkowskian and -- by analytic continuation -- Euclidean space-time. \par
Specifically, we have calculated the $n$-point conformal correlators of the balanced $\mathbb{O},\tilde{\mathbb{O}}$ and unbalanced $\mathbb{S},\bar{\mathbb{S}}$ operators \footnote{In our terminology 'unbalanced' and 'balanced' refers to either the different or equal number of dotted and undotted indices that the aforementioned operators respectively possess in the spinorial representation. Unbalanced operators are referred to as 'asymmetric' in \cite{Beisert:2004fv} and 'anisotropic' in \cite{Robertson:1990bf}.} with collinear twist $2$ \cite{BPS1} in their separate sectors and the $3$-point correlators in the mixed sector as well.\par
We have also reconstructed \cite{BPS1} from the $n$-point conformal correlators in each separate sector the corresponding generating functionals,
$\Gamma_{conf}[j_{\mathbb{O}},j_{\tilde{\mathbb{O}}}]=\Gamma_{conf \, sphere}[j_{\mathbb{O}},j_{\tilde{\mathbb{O}}}]+\Gamma_{conf \,torus}[j_{\mathbb{O}},j_{\tilde{\mathbb{O}}}]$ and $\Gamma_{conf}[j_{\mathbb{S}},j_{\bar{\mathbb{S}}}]=\Gamma_{conf \,sphere}[j_{\mathbb{S}},j_{\bar{\mathbb{S}}}]+\Gamma_{conf \, torus}[j_{\mathbb{S}},j_{\bar{\mathbb{S}}}]$, both in Minkowskian and Euclidean space-time and both in the coordinate and momentum representation, that turn out \cite{BPS1} to have the structure of the logarithm of functional determinants.\par 
We have pointed out \cite{BPS1} that the rationale behind the generating functionals $\Gamma_{conf \, torus}[j_{\mathbb{O}},j_{\tilde{\mathbb{O}}}]$ and $\Gamma_{conf \, torus}[j_{\mathbb{S}},j_{\bar{\mathbb{S}}}]$ being the logarithm of functional determinants is the prediction \cite{Bochicchio:2016toi} that the LO nonplanar contribution to the nonperturbative effective action \cite{Bochicchio:2016toi} should have the structure of the logarithm of a functional determinant that sums the glueball one-loop diagrams.
Hence, in relation to the above computations, the aim of the present paper is fourfold.\par
First, as opposed to the aforementioned nonperturbative argument, since the computation of the $n$-point conformal correlators is by Feynman diagrams \cite{BPS1}, 
we find a perturbative reason as to why the corresponding generating functionals $\Gamma_{conf}[j_{\mathbb{O}},j_{\tilde{\mathbb{O}}}]$ and $\Gamma_{conf}[j_{\mathbb{S}},j_{\bar{\mathbb{S}}}]$ are the logarithm of functional determinants \cite{BPS1}: As all the aforementioned collinear twist-$2$ operators are quadratic in the gauge field in the light-cone gauge to the lowest order of perturbation theory, the functional integral that defines their generating functional $\mathcal{W}_{conf}[J_{\mathcal O}]$ is Gaussian and leads to a functional determinant involving the corresponding sources $J_{\mathcal O}$. \par
Then, once some technical difficulties are properly understood, a tedious but straightforward computation shows that the functional determinants $\Gamma_{conf}[j_{\mathbb{O}},j_{\tilde{\mathbb{O}}}]$ and $\Gamma_{conf}[j_{\mathbb{S}},j_{\bar{\mathbb{S}}}]$ worked out in \cite{BPS1} actually arise from their very definition as Gaussian functional integrals $\mathcal{W}_{conf}[J_{\mathbb{O}},J_{\tilde{\mathbb{O}}}]$ and $\mathcal{W}_{conf}[J_{\mathbb{S}},J_{\bar{\mathbb{S}}}]$ in the present paper.  \par
Second, our new method to compute the generating functional $\mathcal{W}_{conf}[J_{\mathcal O}]$ allows us to produce a vast generalization of our previous computations.
For example, we calculate the generating functional $\mathcal{W}_{conf}[J_{\mathbb{O}},J_{\tilde{\mathbb{O}}},J_{\mathbb{S}},J_{\bar{\mathbb{S}}}]$ in the mixed balanced/unbalanced sector \footnote{In \cite{BPS1} we could not compute the $n$-point correlators with $n > 3$ in the mixed balanced/unbalanced sector by means of Feynman diagrams because of their complexity. As a consequence we could not reconstruct $\Gamma_{conf}[j_{\mathbb{O}},j_{\tilde{\mathbb{O}}},j_{\mathbb{S}},j_{\bar{\mathbb{S}}}]$ from the correlators.} in YM theory, verify by means of it our previous computation of the mixed $3$-point correlators \cite{BPS1} and compute the mixed $4$-point correlators.\par
Besides, the same technique will allow us to compute \cite{BPSN} -- as the logarithm of a functional (super-)determinant -- the generating functional of the lowest-order conformal correlators of the collinear twist-$2$ operators in $\mathcal{N}=1,2,4$ SUSY YM theory, QCD and $\mathcal{N}=1$ SUSY QCD  -- the super-determinant arising in the supersymmetric cases.\par
Remarkably, the very same technique also applies \cite{BPSN} to all the possibly higher-twist operators that are quadratic \cite{Beisert:2004fv} in the elementary fields in the light-cone gauge to the lowest order of perturbation theory in all the above theories.\par
Third, we work out the UV asymptotics of the Euclidean nonpertubative $n$-point correlators in a certain renormalization scheme described below. The computation is based both on our lowest-order calculation of the Euclidean conformal correlators and its improvement by means of the RG.\par
Generally, collinear twist-$2$ operators $\mathcal{O}_{s_i}$ with maximal spin projection $s_i\geq2$ mix under renormalization with total derivatives of the same operators with lower spins, so that an explicit evaluation of the RG-improved asymptotics of their Euclidean $n$-point correlators in the $\overline{MS}$ renormalization scheme would require the asymptotic estimate -- that seems practically impossible -- of the product of the sum of $O(s_i)$ terms for each $\mathcal{O}_{s_i}$ occurring in the $n$-point correlators because of the triangular nature of their renormalized mixing matrix $Z^{(s_i)}$, not to mention that in general $Z^{(s_i)}$ is actually only known to two loops \cite{Belitsky:1998gc,Braun2,Braun3} in perturbation theory and not in its RG-improved form.\par
Recently, a differential-geometric approach to operator mixing in massless QCD-like theories has been proposed \cite{MB1} precisely to overcome the above difficulty. Specifically, it has been determined under which conditions a renormalization scheme exists where each $Z^{(s_i)}$ may be set in a diagonal canonical form that is one-loop exact to all perturbative orders -- according to the nonresonant diagonalizable $\frac{\gamma_0}{\beta_0}$ case (I) \cite{MB1}.
Moreover, it has been verified \cite{Aglietti:2021bem} that the balanced collinear twist-$2$ operators belong to the case (I) and we show in the present paper that this is also the case for the unbalanced ones \footnote{In both cases numerically up to $s=10^4$.}.\par
Remarkably, in such a scheme the needed terms for the asymptotic evaluation of the $n$-point correlators reduce to just one, since in this scheme the operators are multiplicatively renormalizable and in general their correlators do not vanish to the lowest order of perturbation theory. The nonresonant diagonal scheme may also apply \cite{BPSN} to the collinear twist-$2$ operators in QCD and its supersymmetric extensions above. \par
Fourth, we employ the UV asymptotics of the Euclidean $n$-point correlators in the nonresonant diagonal scheme to work out the corresponding generating functional $\mathcal{W}_{asym}^E[J_{\mathcal O},\lambda]$ that also turns out to be the logarithm of a functional determinant.
In fact, after a suitable rescaling of the operators, it may be decomposed for large $N$ into the sum of the generating functionals of the planar $\mathcal{W}^E_{asym \, sphere}[J_{\mathcal O},\lambda]$ and  LO nonplanar asymptotic correlators $\mathcal{W}^E_{asym \, torus}[J_{\mathcal O},\lambda]$, which has the structure of the logarithm of a functional determinant as well.  \par
This is our last -- and perhaps most remarkable -- step. Indeed, according to the prediction in \cite{Bochicchio:2016toi} the nonperturbative glueball one-loop 
generating functional $\mathcal{W}^E_{torus}[J_{\mathcal O},\lambda]$ should have the structure of the logarithm of a functional determinant and as a consequence $\mathcal{W}^E_{asym \, torus}[J_{\mathcal O},\lambda]$ should inherit the very same structure as well. \par
In a forthcoming paper we will further investigate the above nonperturbative interpretation. 
In any case the generating functionals of planar $\mathcal{W}^E_{asym \, sphere}[J_{\mathcal O},\lambda]$ and LO nonplanar asymptotic correlators $\mathcal{W}^E_{asym \, torus}[J_{\mathcal O},\lambda]$ set strong UV asymptotic constraints on the nonperturbative solution of large-$N$ YM theory and may be a pivotal guide for the search of such a solution.

  \section{Plan of the paper} \label{s2}
  
In section \ref{s3} we compute the generating functional $\mathcal{W}_{conf}[J_{\mathcal O}]=\mathcal{W}_{conf}[J_{\mathbb{O}},J_{\tilde{\mathbb{O}}},J_{\mathbb{S}},J_{\bar{\mathbb{S}}}]$ in the coordinate representation of all the correlators of collinear twist-$2$ operators (appendix \ref{B}) to the lowest order of perturbation theory from the YM path integral in Minkowskian space-time.\par
In section \ref{s4} we restrict $\mathcal{W}_{conf}[J_{\mathbb{O}},J_{\tilde{\mathbb{O}}},J_{\mathbb{S}},J_{\bar{\mathbb{S}}}]$ to the generating functionals in the separate balanced and unbalanced sectors, $\mathcal{W}_{conf}[J_{\mathbb{O}},J_{\tilde{\mathbb{O}}}]$ and $\mathcal{W}_{conf}[J_{\mathbb{S}},J_{\bar{\mathbb{S}}}]$, and connect them to the generating functionals, $\Gamma_{conf}[j_{\mathbb{O}},j_{\tilde{\mathbb{O}}}]$ and $\Gamma_{conf}[j_{\mathbb{S}},j_{\bar{\mathbb{S}}}]$, previously computed in \cite{BPS1} directly from the corresponding correlators, finding perfect agreement. \par
In section \ref{momgen} we work out the generating functional $\mathcal{W}_{conf}[J_{\mathcal O}]$ in the momentum representation.\par
In section \ref{s5} we calculate the $3$- and $4$-point correlators in the mixed sector from $\mathcal{W}_{conf}[J_{\mathcal O}]$. The result agrees with our previous computation \cite{BPS1} of the mixed $3$-point correlators. \par
In section \ref{s0} we recall some recent results \cite{MB1} on operator mixing and apply them to work out the UV asymptotics in the coordinate representation of the Euclidean $n$-point correlators of collinear twist-$2$ operators in the aforementioned nonresonant diagonal scheme \cite{MB1}.\par
In section \ref{s6} we compute the generating functional in the coordinate representation of the RG-improved correlators in Euclidean space-time $\mathcal{W}^E_{asym}[J_{\mathcal O},\lambda]=\mathcal{W}^E_{asym \, sphere}[J_{\mathcal O},\lambda]+\mathcal{W}^E_{asym \,torus}[J_{\mathcal O},\lambda]$.\par
In appendix \ref{A} we compute the functional determinant that leads to the generating functional $\mathcal{W}_{conf}[J_{\mathcal O}]$.\par
In appendix \ref{B} we recall the definition of the standard basis \cite{BPS1} of collinear twist-$2$ operators and describe its conformal properties to the leading and next-to-leading perturbative order.

\section{Generating functional $\mathcal{W}_{conf}$ of the Minkowskian conformal correlators from the YM functional integral  \label{s3}}

The Minkowskian YM action is:
  \begin{equation}
  S_{YM} = -\frac{1}{2}\int d^4x\,\Tr F_{\mu\nu}F^{\mu\nu}
  \end{equation}
  where:
  \begin{align}
  F_{\mu\nu} = \partial_\mu A_\nu - \partial_\nu A_\mu +i\frac{g}{\sqrt{N}}\left[A_\mu,A_\nu\right]
  \end{align}
  and:
  \begin{equation}
  A_\mu= A_\mu^a T^a
  \end{equation}
with the Hermitian generators of the $SU(N)$ Lie algebra:
  \begin{equation}
  \left[T^a,T^b\right] = i f^{abc}T^c
  \end{equation}
 in the fundamental representation normalized as: 
  \begin{equation}
  \Tr(T^aT^b) = \frac{\delta^{ab}}{2}
 \end{equation}
 and $g^2 = Ng^2_{YM}$ the 't Hooft coupling \cite{tHooft:1973alw}. We set:
  \begin{align}
  &V_+ = \frac{V_0+ V_3}{\sqrt{2}}\qquad V_{-} = \frac{V_0- V_3}{\sqrt{2}}\nonumber\\
  &V = \frac{V_1+i V_2}{\sqrt{2}}\qquad \bar{V} = \frac{V_1-i V_2}{\sqrt{2}}
  \end{align}
  for a vector $V_{\mu}$.
  We choose the light-cone gauge:
  \begin{equation}
  A_+ = 0
  \end{equation}
After integrating out $A_-$, the YM action in the light-cone gauge reads \cite{Belitsky:2004sc}:
  \begin{align}
S_{YM}(A, \bar A) = -&\int \bar{A}^a \square A^a+2\frac{g}{\sqrt{N}} f^{abc}(A^a\partial_+\bar{A}^b\bar{\partial}\partial_+^{-1}A^c+\bar{A}^a\partial_+A^b\partial\partial_+^{-1}\bar{A}^c) \nonumber\\
&+2\frac{g^2}{N}f^{abc}f^{ade}\partial_+^{-1}(A^b\partial_+ \bar{A}^c)\partial_+^{-1}(\bar{A}^d\partial_+A^e)  \,d^4x
\end{align} 
 The v.e.v. of a product of local gauge-invariant operators $\mathcal{O}_i(A) $ that do not depend on $A_{-}$ reads:
\begin{equation}
  \langle\mathcal{O}_1(x_1)\ldots \mathcal{O}_n(x_n)\rangle =\frac{1}{Z}\int \mathcal{D}A\mathcal{D}\bar{A}\, e^{iS_{YM}(A,\bar A)} \mathcal{O}_1(x_1)\ldots \mathcal{O}_n(x_n)
 \end{equation}
 To the leading perturbative order it reduces to:
  \begin{equation}
  \langle \mathcal{O}_1(x_1)\ldots \mathcal{O}_n(x_n)\rangle =\frac{1}{Z}\int \mathcal{D}A\mathcal{D}\bar{A}\, e^{-i \int d^4x\, \bar{A}^a \square A^a}\mathcal{O}_1(x_1)\ldots \mathcal{O}_n(x_n)
  \end{equation}
 with:
 \begin{equation}
 \square= g^{\mu\nu}\partial_{\mu}\partial_{\nu}=\partial_0^2-\sum_{i=1}^{3}\partial_i^2
 \end{equation}
where we employ the mostly minus metric $g^{\mu\nu}$ in Minkowskian space-time \cite{BPS1}. The corresponding generating functional reads to the leading order: 
  \begin{align}
  \mathcal{Z}_{conf}[J_{\mathcal{O}}] = \frac{1}{Z}\int \mathcal{D}A\mathcal{D}\bar{A}\, e^{-i \int d^4x\, \bar{A}^a \square A^a}\exp\left(\int d^4x\sum_{i}\, J_{\mathcal{O}_i}\mathcal{O}_i\right)
\end{align}

  \subsection{Standard basis}
 
The collinear twist-$2$ operators in the standard basis \cite{BPS1} are (appendix \ref{B}):
  \begin{align}
  \mathbb{O}_{s} &=\frac{1}{2}\bar{A}^a(x) \overleftarrow{\partial}_+ (i\overrightarrow{\partial}_++i\overleftarrow{\partial}_+)^{s-2}C^{\frac{5}{2}}_{s-2}\left(\frac{\overrightarrow{\partial}_+-\overleftarrow{\partial}_+}{\overrightarrow{\partial}_++\overleftarrow{\partial}_+}\right)\overrightarrow{\partial}_+ A^a(x)\nonumber\\
  &=\frac{1}{2}\bar{A}^a(x) \mathcal{Y}_{s-2}^{\frac{5}{2}}(\overrightarrow{\partial}_+,\overleftarrow{\partial}_+) A^a(x)\qquad\qquad s=2,4,6,\ldots\nonumber\\
  \tilde{\mathbb{O}}_{s} &=\frac{1}{2}\bar{A}^a(x) \overleftarrow{\partial}_+ (i\overrightarrow{\partial}_++i\overleftarrow{\partial}_+)^{s-2}C^{\frac{5}{2}}_{s-2}\left(\frac{\overrightarrow{\partial}_+-\overleftarrow{\partial}_+}{\overrightarrow{\partial}_++\overleftarrow{\partial}_+}\right)\overrightarrow{\partial}_+ A^a(x)\nonumber\\
  &=\frac{1}{2}\bar{A}^a(x) \mathcal{H}_{s-2}^{\frac{5}{2}}(\overrightarrow{\partial}_+,\overleftarrow{\partial}_+) A^a(x)\qquad\qquad s=3,5,7,\ldots\nonumber\\
  \mathbb{S}_{s} &=\frac{1}{2\sqrt{2}}\bar{A}^a(x) \overleftarrow{\partial}_+ (i\overrightarrow{\partial}_++i\overleftarrow{\partial}_+)^{s-2}C^{\frac{5}{2}}_{s-2}\left(\frac{\overrightarrow{\partial}_+-\overleftarrow{\partial}_+}{\overrightarrow{\partial}_++\overleftarrow{\partial}_+}\right)\overrightarrow{\partial}_+ \bar{A}^a(x)\nonumber\\
  &=\frac{1}{2\sqrt{2}}\bar{A}^a(x) \mathcal{Y}_{s-2}^{\frac{5}{2}}(\overrightarrow{\partial}_+,\overleftarrow{\partial}_+) \bar{A}^a(x)\qquad\qquad s=2,4,6,\ldots\nonumber\\
  \bar{\mathbb{S}}_{s} &=\frac{1}{2\sqrt{2}}{A}^a(x) \overleftarrow{\partial}_+ (i\overrightarrow{\partial}_++i\overleftarrow{\partial}_+)^{s-2}C^{\frac{5}{2}}_{s-2}\left(\frac{\overrightarrow{\partial}_+-\overleftarrow{\partial}_+}{\overrightarrow{\partial}_++\overleftarrow{\partial}_+}\right)\overrightarrow{\partial}_+ {A}^a(x)\nonumber\\
  &=\frac{1}{2\sqrt{2}}{A}^a(x) \mathcal{Y}_{s-2}^{\frac{5}{2}}(\overrightarrow{\partial}_+,\overleftarrow{\partial}_+) {A}^a(x)\qquad\qquad s=2,4,6,\ldots
  \end{align}
  where the sum over repeated color indices is understood, and \cite{BPS1}:
  \begin{align}
  \label{defY}
  \mathcal{Y}_{s-2}^{\frac{5}{2}}(\overrightarrow{\partial}_+,\overleftarrow{\partial}_+) &= \overleftarrow{\partial}_+ (i\overrightarrow{\partial}_++i\overleftarrow{\partial}_+)^{s-2}C^{\frac{5}{2}}_{s-2}\left(\frac{\overrightarrow{\partial}_+-\overleftarrow{\partial}_+}{\overrightarrow{\partial}_++\overleftarrow{\partial}_+}\right)\overrightarrow{\partial}_+\nonumber\\
  &=\frac{\Gamma(3)\Gamma(s+3)}{\Gamma(5)\Gamma(s+1)}i^{s-2}\sum_{k=0}^{s-2} {s\choose k}{s\choose k+2}(-1)^{s-k} \overleftarrow{\partial}_{+}^{s-k-1} \overrightarrow{\partial}_{+}^{k+1}
  \end{align}
 for even spin, and: 
  \begin{align}
	\label{defH}
	\mathcal{H}_{s-2}^{\frac{5}{2}}(\overrightarrow{\partial}_+,\overleftarrow{\partial}_+) &= \overleftarrow{\partial}_+ (i\overrightarrow{\partial}_++i\overleftarrow{\partial}_+)^{s-2}C^{\frac{5}{2}}_{s-2}\left(\frac{\overrightarrow{\partial}_+-\overleftarrow{\partial}_+}{\overrightarrow{\partial}_++\overleftarrow{\partial}_+}\right)\overrightarrow{\partial}_+\nonumber\\
	&=\frac{\Gamma(3)\Gamma(s+3)}{\Gamma(5)\Gamma(s+1)}i^{s-2}\sum_{k=0}^{s-2} {s\choose k}{s\choose k+2}(-1)^{s-k} \overleftarrow{\partial}_{+}^{s-k-1} \overrightarrow{\partial}_{+}^{k+1}
\end{align}
for odd spin. The corresponding generating functional of conformal correlators to the leading order reads:
  \begin{align}
  &\mathcal{Z}_{conf}[J_{\mathbb{O}},J_{\tilde{\mathbb{O}}},J_{\mathbb{S}},J_{\bar{\mathbb{S}}}]\nonumber\\
   &=\frac{1}{Z} \int \mathcal{D}A\mathcal{D}\bar{A}\, e^{-i \int d^4x\, \bar{A}^a \square A^a}\exp\left(\int d^4x\sum_s\, J_{\mathbb{O}_s}\mathbb{O}_s+J_{\tilde{\mathbb{O}}_s}\tilde{\mathbb{O}}_s+J_{\mathbb{S}_s}\mathbb{S}_s+J_{\bar{\mathbb{S}}_s}\bar{\mathbb{S}}_s\right)
  \end{align}
where the currents $J_{\mathbb{O}_s}, J_{\tilde{\mathbb{O}}_s}, J_{\mathbb{S}_s}, J_{\bar{\mathbb{S}}_s}$ are defined to be zero for $s$ different from the spin of the corresponding operator. Explicitly:
\begin{align}
  &\mathcal{Z}_{conf}[J_{\mathbb{O}},J_{\tilde{\mathbb{O}}},J_{\mathbb{S}},J_{\bar{\mathbb{S}}}] \nonumber\\
  &= \frac{1}{Z}\int \mathcal{D}A\mathcal{D}\bar{A}\, e^{-i\int d^4x\, \bar{A}^a \square A^a}\exp\Big(\frac{1}{2}\int d^4x\sum_s\,J_{\mathbb{O}_{s}}\bar{A}^a(x) \mathcal{Y}_{s-2}^{\frac{5}{2}}(\overrightarrow{\partial}_+,\overleftarrow{\partial}_+) A^a(x)
  \nonumber\\
  &\quad+J_{\tilde{\mathbb{O}}_{s}}\bar{A}^a(x) \mathcal{H}_{s-2}^{\frac{5}{2}}(\overrightarrow{\partial}_+,\overleftarrow{\partial}_+) A^a(x)\nonumber\\
  &\quad+J_{\mathbb{S}_{s}}\frac{1}{\sqrt{2}}\bar{A}^a(x) \mathcal{Y}_{s-2}^{\frac{5}{2}}(\overrightarrow{\partial}_+,\overleftarrow{\partial}_+) \bar{A}^a(x)+J_{\bar{\mathbb{S}}_{s}}\frac{1}{\sqrt{2}}{A}^a(x) \mathcal{Y}_{s-2}^{\frac{5}{2}}(\overrightarrow{\partial}_+,\overleftarrow{\partial}_+) {A}^a(x)\Big)
  \end{align}
The above functional integral is quadratic in the elementary fields and therefore it may be computed exactly. Employing the symmetry properties of the Gegenbauer polynomials (appendix \ref{B}):
 \begin{align}
&\mathbb{O}_{s}=\frac{1}{4}\left(\bar{A}^a(x) \mathcal{Y}_{s-2}^{\frac{5}{2}}(\overrightarrow{\partial}_+,\overleftarrow{\partial}_+) A^a(x)+A^a(x) \mathcal{Y}_{s-2}^{\frac{5}{2}}(\overrightarrow{\partial}_+,\overleftarrow{\partial}_+) \bar{A}^a(x)\right)\nonumber\\
&\tilde{\mathbb{O}}_{s} = \frac{1}{4}\left(\bar{A}^a(x) \mathcal{H}_{s-2}^{\frac{5}{2}}(\overrightarrow{\partial}_+,\overleftarrow{\partial}_+) A^a(x)-A^a(x) \mathcal{H}_{s-2}^{\frac{5}{2}}(\overrightarrow{\partial}_+,\overleftarrow{\partial}_+) \bar{A}^a(x)\right)
\end{align}
 we obtain: 
    \begin{align}
  \mathcal{Z}_{conf}[J_{\mathbb{O}},J_{\tilde{\mathbb{O}}},J_{\mathbb{S}},J_{\bar{\mathbb{S}}}] =& \frac{1}{Z} \int \mathcal{D}A\mathcal{D}\bar{A}\, e^{-\frac{1}{2}\int d^4x\, \begin{pmatrix}
  	\bar{A}^a(x)&A^a(x)
  	\end{pmatrix} \mathcal{M}^{ab}  \begin{pmatrix}
  	 &A^b(x) &\\
  	 &\bar{A}^b(x) & 
  	\end{pmatrix}}
  \end{align}
  with:
 \begin{align}
	\label{matrixM}
	&\mathcal{M}^{ab}=\delta^{ab}\scalebox{0.85}{$\begin{pmatrix}i\square-\frac{1}{2}\sum_sJ_{\mathbb{O}_{s}}\otimes\mathcal{Y}_{s-2}^{\frac{5}{2}}-\frac{1}{2}\sum_sJ_{\tilde{\mathbb{O}}_{s}}\otimes\mathcal{H}_{s-2}^{\frac{5}{2}} & -\frac{1}{\sqrt{2}}\sum_s J_{\mathbb{S}_{s}}\otimes\mathcal{Y}_{s-2}^{\frac{5}{2}}\\ -\frac{1}{\sqrt{2}}\sum_sJ_{\bar{\mathbb{S}}_{s}}\otimes\mathcal{Y}_{s-2}^{\frac{5}{2}}&i\square-\frac{1}{2}\sum_sJ_{\mathbb{O}_{s}}\otimes\mathcal{Y}_{s-2}^{\frac{5}{2}}+\frac{1}{2}\sum_sJ_{\tilde{\mathbb{O}}_{s}}\otimes\mathcal{H}_{s-2}^{\frac{5}{2}}  \end{pmatrix}$}
\end{align}
where we have introduced the symbol $\otimes$ to imply that the right and left derivatives do not act on the sources $J$.
Hence, performing the Gaussian integral we get:
  \begin{equation}
\mathcal{Z}_{conf}[J_{\mathbb{O}},J_{\tilde{\mathbb{O}}},J_{\mathbb{S}},J_{\bar{\mathbb{S}}}] =\Det^{-\frac{1}{2}}( \mathcal{M})
\end{equation}
where the above determinant -- up to a trivial normalization factor -- is computed in eq. \eqref{detMat}. Hence, the generating functional of the connected correlators reads:
\begin{align}
	  	\label{Wgen1}
	&\mathcal{W}_{conf}[J_{\mathbb{O}},J_{\tilde{\mathbb{O}}},J_{\mathbb{S}},J_{\bar{\mathbb{S}}}]  = 	\log\mathcal{Z}_{conf}[J_{\mathbb{O}},J_{\tilde{\mathbb{O}}},J_{\mathbb{S}},J_{\bar{\mathbb{S}}}] \nonumber\\ &=-\frac{1}{2}\log\Det\left(\mathcal{I}+\frac{1}{2}i\square^{-1}J_{\mathbb{O}_{s}}\otimes\mathcal{Y}_{s-2}^{\frac{5}{2}}+\frac{1}{2}i\square^{-1}J_{\tilde{\mathbb{O}}_{s}}\otimes\mathcal{H}_{s-2}^{\frac{5}{2}} \right)\nonumber\\
	&\quad-\frac{1}{2}\log\Det\Big(\mathcal{I}+\frac{1}{2}i\square^{-1}J_{\mathbb{O}_{s}}\otimes\mathcal{Y}_{s-2}^{\frac{5}{2}}-\frac{1}{2}i\square^{-1}J_{\tilde{\mathbb{O}}_{s}}\otimes\mathcal{H}_{s-2}^{\frac{5}{2}}\Big)\nonumber\\
	&\quad-\frac{1}{2}\log\Det\Bigg[\mathcal{I}-\frac{1}{2}\left(\mathcal{I}+\frac{1}{2}i\square^{-1}J_{\mathbb{O}_{s}}\otimes\mathcal{Y}_{s-2}^{\frac{5}{2}}-\frac{1}{2}i\square^{-1}J_{\tilde{\mathbb{O}}_{s}}\otimes\mathcal{H}_{s-2}^{\frac{5}{2}}\right)^{-1}\nonumber\\
	&\quad\quad i\square^{-1}J_{\bar{\mathbb{S}}_{s_1}} \otimes\mathcal{Y}_{s_1-2}^{\frac{5}{2}}\left(\mathcal{I}+\frac{1}{2}i\square^{-1}J_{\mathbb{O}_{s_2}}\otimes\mathcal{Y}_{s_2-2}^{\frac{5}{2}}+\frac{1}{2}i\square^{-1}J_{\tilde{\mathbb{O}}_{s_2}}\otimes\mathcal{H}_{s_2-2}^{\frac{5}{2}} \right)^{-1}\nonumber\\
	&\quad\quad i\square^{-1}J_{\mathbb{S}_{s_3}} \otimes\mathcal{Y}_{s_3-2}^{\frac{5}{2}} \Bigg]
\end{align}
where $\mathcal{I}$ is the identity in both color and space-time and the sum over repeated spin indices is understood. 
After rescaling the operators:
\begin{equation}
	\label{rescale}
	\mathcal{O}'_s(x)=\frac{1}{N}\frac{2\Gamma(5)\Gamma(s+1)}{\Gamma(3)\Gamma(s+3)}\mathcal{O}_s(x)
\end{equation}
so that their $2$-point correlators are of order $1$ for large $N$, we obtain more explicitly:
  \begin{align}
  	\label{wexplicit}
	&{\mathcal{W}_{conf}}[J_{\mathbb{O}'},J_{\tilde{\mathbb{O}}'},J_{\mathbb{S}'},J_{\bar{\mathbb{S}}'}]=\nonumber\\
	& -\frac{N^2-1}{2}\log\Det\left(I+\frac{1}{N}\sum_{k=0}^{s-2}{s\choose k}{s\choose k+2}(i\overrightarrow{\partial}_+)^{s-k-1}i \square^{-1}(J_{\mathbb{O}'_{s}}+J_{\tilde{\mathbb{O}}'_{s}})(i\overrightarrow{\partial}_+)^{k+1} \right)\nonumber\\
	&-\frac{N^2-1}{2}\log\Det\left(I+\frac{1}{N}\sum_{k=0}^{s-2}{s\choose k}{s\choose k+2}(i\overrightarrow{\partial}_+)^{s-k-1}i \square^{-1}(J_{\mathbb{O}'_{s}}-J_{\tilde{\mathbb{O}}'_{s}})(i\overrightarrow{\partial}_+)^{k+1} \right)\nonumber\\
	&-\frac{N^2-1}{2}\log\Det\Bigg[I-\frac{2}{N^2}\left(I+\frac{1}{N}\sum_{k=0}^{s-2}{s\choose k}{s\choose k+2}(i\overrightarrow{\partial}_+)^{s-k-1}i \square^{-1}(J_{\mathbb{O}'_{s}}-J_{\tilde{\mathbb{O}}'_{s}})(i\overrightarrow{\partial}_+)^{k+1} \right)^{-1}\nonumber\\
	& \quad\sum_{k_1=0}^{s_1-2}{s_1\choose k_1}{s_1\choose k_1+2}(i\overrightarrow{\partial}_+)^{s_1-k_1-1}i \square^{-1}J_{\bar{\mathbb{S}}'_{s_1}}(i\overrightarrow{\partial}_+)^{k_1+1} \nonumber\\
	&\quad\left(I+\frac{1}{N}\sum_{k_2=0}^{s_2-2}{s_2\choose k_2}{s_2\choose k_2+2}(i\overrightarrow{\partial}_+)^{s_2-k_2-1}i \square^{-1}(J_{\mathbb{O}'_{s_2}}+J_{\tilde{\mathbb{O}}'_{s_2}})(i\overrightarrow{\partial}_+)^{k_2+1} \right)^{-1}\nonumber\\
	& \quad\sum_{k_3=0}^{s_3-2}{s_3\choose k_3}{s_3\choose k_3+2}(i\overrightarrow{\partial}_+)^{s_3-k_3-1}i \square^{-1}J_{\mathbb{S}'_{s_3}}(i\overrightarrow{\partial}_+)^{k_3+1}\Bigg]
\end{align}
where $I$ is the identity in space-time and we have already performed the color trace. To obtain eq. \eqref{wexplicit} we have employed the definitions in eqs. \eqref{defY} and \eqref{defH}, and:
\begin{align}
i\square^{-1}\overleftarrow{\partial}_{+}^{s-k-1} = (-1)^{s-k-1}\overrightarrow{\partial}_{+}^{s-k-1} i\square^{-1}
\end{align}
that follows from (minus) the propagator in the coordinate representation \cite{BPS1}:
\begin{equation}
	\label{propagator}
	\frac{1}{4\pi^2}\frac{1}{\rvert x-y \rvert^2-i\epsilon} =i \square^{-1}(x-y)
\end{equation}

 \section{Connection of $\mathcal{W}_{conf}$ with the generating functionals $\Gamma_{conf}$ \label{s4}}
 
Remarkably, ${\mathcal{W}_{conf}}[J_{\mathbb{O}},J_{\tilde{\mathbb{O}}},J_{\mathbb{S}},J_{\bar{\mathbb{S}}}]$ is the generating functional of all the Minkowskian conformal correlators that extends the generating functionals in the separate balanced and unbalanced sectors obtained \cite{BPS1} by guessing their structure from the perturbative computation of the corresponding $n$-point correlators.
However, even restricting to the separate balanced and unbalanced sectors, the connection with the approach in \cite{BPS1} is not immediately visible.
Indeed, the connected generating functionals in the separate sectors read \cite{BPS1}:
 \begin{align} \label{4.1}
 \nonumber
 \Gamma_{conf}[j_{\mathbb{O}},j_{\tilde{\mathbb{O}}}] =& -\frac{N^2-1}{2}\log \Det\left(\mathbb{I}+\mathcal{D}^{-1}j_{\mathbb{O}}+\mathcal{D}^{-1}j_{\tilde{\mathbb{O}}}\right)\\\nonumber
 &-\frac{N^2-1}{2}\log \Det\left(\mathbb{I}+\mathcal{D}^{-1}j_{\mathbb{O}}-\mathcal{D}^{-1}j_{\tilde{\mathbb{O}}}\right)\\\nonumber
\nonumber\\
 \Gamma_{conf}[j_{\mathbb{S}},j_{\bar{\mathbb{S}}}]=& -\frac{N^2-1}{2}\log \Det\left(\mathbb{I}-2\mathcal{D}^{-1}j_{\bar{\mathbb{S}}}\mathcal{D}^{-1}j_{\mathbb{S}}\right)
 \end{align}
in Minkowskian space-time, where $\mathbb{I}$ is the identity in both space-time and the discrete indices defined below.
By making manifest the space-time and discrete-indices dependence in the kernels of the integral operators above, eq. \eqref{4.1} reads in the coordinate representation \cite{BPS1}:
 \begin{align}
 \nonumber\
& \Gamma_{conf}[j_{\mathbb{O}},j_{\tilde{\mathbb{O}}}]\nonumber\\
 &=-\frac{N^2-1}{2}\log \Det\Big(\delta_{s_1k_1,s_2k_2}\delta^{(4)}(x-y)+\mathcal{D}^{-1}_{s_1k_1,s_2k_2}(x-y) (j_{\mathbb{O}_{s_2k_2}}(y)+j_{\tilde{\mathbb{O}}_{s_2k_2}}(y)) \Big)\nonumber\\
 &\quad-\frac{N^2-1}{2}\log \Det\Big(\delta_{s_1k_1,s_2k_2}\delta^{(4)}(x-y)+\mathcal{D}^{-1}_{s_1k_1,s_2k_2}(x-y) (j_{\mathbb{O}_{s_2k_2}}(y)-j_{\tilde{\mathbb{O}}_{s_2k_2}}(y)) \Big)\nonumber\\\nonumber\\
& \Gamma_{conf}[j_{\mathbb{S}},j_{\bar{\mathbb{S}}}] \nonumber\\
&=-\frac{N^2-1}{2}\log \Det\Big(\delta_{s_1k_1,s_2k_2}\delta^{(4)}(x-y)\nonumber\\
&\quad-2\int d^4z\sum_{sk}\mathcal{D}^{-1}_{s_1k_1,sk}(x-z)j_{\bar{\mathbb{S}}_{sk}}(z)\mathcal{D}^{-1}_{sk,s_2k_2}(z-y)j_{\mathbb{S}_{s_2k_2}}(y)\Big)
 \end{align}
where, by a slight abuse of notation, we have displayed as arguments of the determinants the corresponding integral kernels, with \cite{BPS1}:
 \begin{align}
 \label{kernelD}
 \mathcal{D}^{-1}_{s_1k_1,s_2k_2}(x-y) &=\frac{i^{s_1+1}}{2}\frac{\Gamma(3)\Gamma(s_1+3)}{\Gamma(5)\Gamma(s_1+1)}{s_1\choose k_1}{s_2\choose k_2+2}(-\partial_{+})^{s_1-k_1+k_2}\square^{-1}(x-y)\nonumber\\ &=\frac{i^{s_1}}{8\pi^2}\frac{\Gamma(3)\Gamma(s_1+3)}{\Gamma(5)\Gamma(s_1+1)}{s_1\choose k_1}{s_2\choose k_2+2}(-\partial_{+})^{s_1-k_1+k_2}\frac{1}{\rvert x-y\rvert^2-i\epsilon}
 \end{align}
 Equivalently, we may employ the modified kernel:
  \begin{align}
 \label{kernelD'}
 \mathcal{D'}^{-1}_{s_1k_1,s_2k_2}(x-y) &=\frac{i}{2}\frac{\Gamma(3)\Gamma(s_1+3)}{\Gamma(5)\Gamma(s_1+1)}{s_1\choose k_1}{s_2\choose k_2+2}(-i\partial_{+})^{s_1-k_1+k_2}\square^{-1}(x-y)\nonumber\\ &=\frac{1}{8\pi^2}\frac{\Gamma(3)\Gamma(s_1+3)}{\Gamma(5)\Gamma(s_1+1)}{s_1\choose k_1}{s_2\choose k_2+2}(-i\partial_{+})^{s_1-k_1+k_2}\frac{1}{\rvert x-y\rvert^2-i\epsilon}
 \end{align}
It gives origin to the same correlators, since it differs from $\mathcal{D}^{-1}_{s_1k_1,s_2k_2}(x-y)$ by the factor $i^{-k_1+k_2}$ that cancels in the loops.\par
 The currents $j_{\mathcal{O}_{sk}}$ are dual to the component operators $\mathcal{O}_{sk}$ that are employed to construct the conformal operators $\mathcal{O}_s$ \cite{BPS1}:
 \begin{equation}
 	\mathcal{O}_s = \sum_{k=0}^{l}\mathcal{O}_{sk}
 \end{equation}
with $l=s-2$ for the standard basis.
Consequently, the $n$-point correlators satisfy \cite{BPS1}:
 \begin{align}
 	\label{gammacorr}
 	\langle \mathcal{O}_{s_1}(x_1)\ldots\mathcal{O}_{s_n}(x_n)\rangle =\sum_{k_1 = 0}^{l_1} \frac{\delta}{\delta j_{\mathcal{O}_{s_1k_1}}(x_1)}\cdots\sum_{k_n = 0}^{l_n}\frac{\delta}{\delta j_{\mathcal{O}_{s_nk_n}}(x_n)} {\Gamma_{conf}}[j_{\mathcal{O}}]
 \end{align}
 Hence, we should demonstrate that:
 \begin{equation}
 	\label{dictionary}
 	\sum_{k_1 = 0}^{l_1} \frac{\delta}{\delta j_{\mathcal{O}_{s_1k_1}}(x_1)}\cdots\sum_{k_n = 0}^{l_n}\frac{\delta}{\delta j_{\mathcal{O}_{s_nk_n}}(x_n)} {\Gamma_{conf}}[j_{\mathcal{O}}] = \frac{\delta}{\delta J_{\mathcal{O}_{s_1}}(x_1)}\cdots\frac{\delta}{\delta J_{\mathcal{O}_{s_n}}(x_n)} {\mathcal{W}_{conf}}[J_{\mathcal{O}}]
 \end{equation}
with the functional derivatives computed at $j_{\mathcal{O}_{sk}}=0$ and $J_{\mathcal{O}_{s}}=0$ respectively.
While ${\Gamma_{conf}}[j_{\mathcal{O}}]$ involves determinants of integral operators formally of Fredholm type, the path-integral computation of ${\mathcal{W}_{conf}}[J_{\mathcal{O}}]$ involves determinants of quadratic forms that originally arise from their very definition in eq. \eqref{Wgen1} by employing both left and right derivatives in eqs. \eqref{defY} and \eqref{defH}.
This is the source of some technical complications that we resolve momentarily creating a dictionary that relates ${\mathcal{W}_{conf}}[J_{\mathcal{O}}]$ to ${\Gamma_{conf}}[j_{\mathcal{O}}]$.\par

\subsection{An example of the dictionary}

We choose as example the generating functional ${\mathcal{W}_{conf}}[J_{\mathbb{O}},0,0,0]$ restricted to the balanced operators with even spin in the standard basis. We formally expand the logarithm of the functional determinant: 
\begin{align}
	&{\mathcal{W}_{conf}}[J_{\mathbb{O}},0,0,0] \nonumber\\
	=&(N^2-1)\sum_{l=1}^{\infty}\frac{(-1)^l}{2^ll}\int d^4x_1\ldots d^4x_l\,\sum_{s_1k_1}\ldots \sum_{s_lk_l}{s_1\choose k_1}{s_1\choose k_1+2}\ldots {s_l\choose k_l}{s_l\choose k_l+2}\nonumber\\
	&\frac{\Gamma(3)\Gamma(s_1+3)}{\Gamma(5)\Gamma(s_1+1)}\cdots \frac{\Gamma(3)\Gamma(s_l+3)}{\Gamma(5)\Gamma(s_l+1)}\nonumber\\
	&(i\overrightarrow{\partial}_{x_1^+})^{s_1-k_1-1}i\square^{-1}(x_1-x_2)J_{\mathbb{O}_{s_1}}(x_2)(i\overrightarrow{\partial}_{x_2^+})^{k_1+1}\nonumber\\
	&(i\overrightarrow{\partial}_{x_2^+})^{s_2-k_2-1}i\square^{-1}(x_2-x_3)J_{\mathbb{O}_{s_2}}(x_3)(i\overrightarrow{\partial}_{x_3^+})^{k_2+1}\nonumber\\
	&\ldots(i\overrightarrow{\partial}_{x_l^+})^{s_l-k_l-1}i\square^{-1}(x_l-x_1)J_{\mathbb{O}_{s_l}}(x_1)(i\overrightarrow{\partial}_{x_1^+})^{k_l+1}
\end{align}
and we combine together the derivatives with respect to the same coordinate:
\begin{align}
	\label{w4}
	&{\mathcal{W}_{conf}}[J_{\mathbb{O}},0,0,0] \nonumber\\
	=&(N^2-1)\sum_{l=1}^{\infty}\frac{(-1)^l}{2^ll}\int d^4x_1\ldots d^4x_l\,\sum_{s_1k_1}\ldots \sum_{s_lk_l}{s_1\choose k_1}{s_1\choose k_1+2}\ldots {s_l\choose k_l}{s_l\choose k_l+2}\nonumber\\
	&\frac{\Gamma(3)\Gamma(s_1+3)}{\Gamma(5)\Gamma(s_1+1)}\cdots \frac{\Gamma(3)\Gamma(s_l+3)}{\Gamma(5)\Gamma(s_l+1)}\nonumber\\
	&(i\overrightarrow{\partial}_{x_1^+})^{s_1-k_1+k_l}i\square^{-1}(x_1-x_2)J_{\mathbb{O}_{s_1}}(x_2)(i\overrightarrow{\partial}_{x_2^+})^{s_2-k_2+k_1}i\square^{-1}(x_2-x_3)J_{\mathbb{O}_{s_2}}(x_3)\nonumber\\
	&\ldots(i\overrightarrow{\partial}_{x_l^+})^{s_l-k_l+k_{l-1}}i\square^{-1}(x_l-x_1)J_{\mathbb{O}_{s_l}}(x_1)
\end{align}
We redefine the spin labels that we sum over, $s_1k_1\rightarrow s_2k_2$, $s_2k_2\rightarrow s_3k_3$, ..., $s_{l-1}k_{l-1}\rightarrow s_l k_l$, $s_{l}k_l\rightarrow s_1 k_1$, to get:
\begin{align}
	&{\mathcal{W}_{conf}}[J_{\mathbb{O}},0,0,0] \nonumber\\
	=&(N^2-1)\sum_{l=1}^{\infty}\frac{(-1)^l}{2^ll}\int d^4x_1\ldots d^4x_l\,\sum_{s_1k_1}\ldots \sum_{s_lk_l}{s_1\choose k_1}{s_1\choose k_1+2}\ldots {s_l\choose k_l}{s_l\choose k_l+2}\nonumber\\
	&\frac{\Gamma(3)\Gamma(s_1+3)}{\Gamma(5)\Gamma(s_1+1)}\cdots \frac{\Gamma(3)\Gamma(s_l+3)}{\Gamma(5)\Gamma(s_l+1)}\nonumber\\
	&(i\overrightarrow{\partial}_{x_1^+})^{s_2-k_2+k_1}i\square^{-1}(x_1-x_2)J_{\mathbb{O}_{s_2}}(x_2)(i\overrightarrow{\partial}_{x_2^+})^{s_3-k_3+k_2}i\square^{-1}(x_2-x_3)J_{\mathbb{O}_{s_3}}(x_3)\nonumber\\
	&\ldots(i\overrightarrow{\partial}_{x_l^+})^{s_1-k_1+k_l}i\square^{-1}(x_l-x_1)J_{\mathbb{O}_{s_1}}(x_1)
\end{align}
Further redefining the labels of both the spin and coordinate, $x_{i}s_{i}k_{i}\rightarrow x_{l-i+2}s_{l-i+2}k_{l-i+2}$ for $2 \leq i \leq l$, keeping
$x_1s_1k_1$ fixed, we obtain:
\begin{align}
	&{\mathcal{W}_{conf}}[J_{\mathbb{O}},0,0,0] \nonumber\\
	=&(N^2-1)\sum_{l=1}^{\infty}\frac{(-1)^l}{2^ll}\int d^4x_1\ldots d^4x_l\,\sum_{s_1k_1}\ldots \sum_{s_lk_l}{s_1\choose k_1}{s_1\choose k_1+2}\ldots {s_l\choose k_l}{s_l\choose k_l+2}\nonumber\\
	&\frac{\Gamma(3)\Gamma(s_1+3)}{\Gamma(5)\Gamma(s_1+1)}\cdots \frac{\Gamma(3)\Gamma(s_l+3)}{\Gamma(5)\Gamma(s_l+1)}\nonumber\\
	&(i\overrightarrow{\partial}_{x_1^+})^{s_l-k_l+k_1}i\square^{-1}(x_1-x_l)J_{\mathbb{O}_{s_l}}(x_l)(i\overrightarrow{\partial}_{x_l^+})^{s_{l-1}-k_{l-1}+k_l}i\square^{-1}(x_l-x_{l-1})J_{\mathbb{O}_{s_{l-1}}}(x_{l-1})\nonumber\\
	&\ldots(i\overrightarrow{\partial}_{x_3^+})^{s_2-k_2+k_3}i\square^{-1}(x_3-x_2)J_{\mathbb{O}_{s_2}}(x_2) (i\overrightarrow{\partial}_{x_2^+})^{s_1-k_1+k_2}i\square^{-1}(x_2-x_1)J_{\mathbb{O}_{s_1}}(x_1)
\end{align}
Employing:
\begin{align}
	{\partial}_{x^+_{a}}^{s_{a}-k_{a}+k_b}i\square^{-1}(x_{a}-x_{b})&=(-1)^{s_{a}-k_{a}+k_b} {\partial}_{x^+_{b}}^{s_{a}-k_{a}+k_b} i\square^{-1}(x_{a}-x_{b}) \nonumber \\
	&= (-{\partial}_{x^+_{b}})^{s_{a}-k_{a}+k_b} i\square^{-1}(x_{b}-x_{a}) 
\end{align}
we get:
\begin{align}
	&{\mathcal{W}_{conf}}[J_{\mathbb{O}},0,0,0] \nonumber\\
	=&(N^2-1)\sum_{l=1}^{\infty}\frac{(-1)^l}{2^ll}\int d^4x_1\ldots d^4x_l\,\sum_{s_1k_1}\ldots \sum_{s_lk_l}{s_1\choose k_1}{s_1\choose k_1+2}\ldots {s_l\choose k_l}{s_l\choose k_l+2}\nonumber\\
	&\frac{\Gamma(3)\Gamma(s_1+3)}{\Gamma(5)\Gamma(s_1+1)}\cdots \frac{\Gamma(3)\Gamma(s_l+3)}{\Gamma(5)\Gamma(s_l+1)}\nonumber\\
	&(-i\overrightarrow{\partial}_{x_l^+})^{s_l-k_l+k_1}i\square^{-1}(x_l-x_1)J_{\mathbb{O}_{s_l}}(x_l)(-i\overrightarrow{\partial}_{x_{l-1}^+})^{s_{l-1}-k_{l-1}+k_l}i\square^{-1}(x_{l-1}-x_{l})J_{\mathbb{O}_{s_{l-1}}}(x_{l-1})\nonumber\\
	&\ldots (-i\overrightarrow{\partial}_{x_2^+})^{s_2-k_2+k_3}i\square^{-1}(x_2-x_3)J_{\mathbb{O}_{s_2}}(x_2)(-i\overrightarrow{\partial}_{x_1^+})^{s_1-k_1+k_2}i\square^{-1}(x_1-x_2)J_{\mathbb{O}_{s_1}}(x_1)
\end{align}
Rearranging the position of the various factors we finally obtain:
\begin{align} \label{W}
	&{\mathcal{W}_{conf}}[J_{\mathbb{O}},0,0,0] \nonumber\\
	=&(N^2-1)\sum_{l=1}^{\infty}\frac{(-1)^l}{2^ll}\int d^4x_1\ldots d^4x_l\,\sum_{s_1k_1}\ldots \sum_{s_lk_l}{s_1\choose k_1}{s_1\choose k_1+2}\ldots {s_l\choose k_l}{s_l\choose k_l+2}\nonumber\\
	&\frac{\Gamma(3)\Gamma(s_1+3)}{\Gamma(5)\Gamma(s_1+1)}\cdots \frac{\Gamma(3)\Gamma(s_l+3)}{\Gamma(5)\Gamma(s_l+1)}\nonumber\\
	&(-i\overrightarrow{\partial}_{x_1^+})^{s_1-k_1+k_2}i\square^{-1}(x_1-x_2)J_{\mathbb{O}_{s_2}}(x_2)(-i\overrightarrow{\partial}_{x_2^+})^{s_2-k_2+k_3}i\square^{-1}(x_2-x_3)J_{\mathbb{O}_{s_3}}(x_3)\nonumber\\
	&\ldots(-i\overrightarrow{\partial}_{x_{l-1}^+})^{s_{l-1}-k_{l-1}+k_l}i\square^{-1}(x_{l-1}-x_{l})J_{\mathbb{O}_{s_{l}}}(x_{l}) (-i\overrightarrow{\partial}_{x_l^+})^{s_l-k_l+k_1}i\square^{-1}(x_l-x_1) J_{\mathbb{O}_{s_1}}(x_1)
\end{align}
Given the corresponding expansion for $\Gamma_{conf}[j_\mathbb{O},0]$:
\begin{align} \label{Gamma}
	&\Gamma_{conf}[j_\mathbb{O},0] \nonumber\\
	&=(N^2-1)\sum_{l=1}^{\infty}\frac{(-1)^{l}}{2^ll}\int d^4x_1\ldots d^4x_n\, \sum_{s_1k_1}\ldots\sum_{s_lk_l}{s_1\choose k_1}{s_1\choose k_1+2}\ldots {s_l\choose k_l}{s_l\choose k_l+2}\nonumber\\
	&\frac{\Gamma(3)\Gamma(s_1+3)}{\Gamma(5)\Gamma(s_1+1)}\cdots \frac{\Gamma(3)\Gamma(s_l+3)}{\Gamma(5)\Gamma(s_l+1)}\nonumber\\
	&(-i\partial_{x_1^+})^{s_1-k_1+k_2}i\square^{-1}(x_1-x_2)j_{\mathbb{O}_{s_2k_2}}(x_2)(-i\partial_{x_2^+})^{s_2-k_2+k_3}i\square^{-1}(x_2-x_3)j_{\mathbb{O}_{s_3k_3}}(x_3)\nonumber\\
	&\ldots (-i\partial_{x_{l-1}^+})^{s_{l-1}-k_{l-1}+k_l}i\square^{-1}(x_{l-1}-x_l)j_{\mathbb{O}_{s_lk_l}}(x_l)(-i\partial_{x_l^+})^{s_l-k_l+k_1}i\square^{-1}(x_l-x_1)j_{\mathbb{O}_{s_1k_1}}(x_1)
\end{align}
it is clear that the functional derivatives of the two objects in eqs. \eqref{W} and \eqref{Gamma} yield the very same result, so that eq. \eqref{dictionary} is proved.

\subsection{Dictionary in the general case}

The dictionary in the general case is constructed as follows. We expand the traces of the logarithms in eq. \eqref{Wgen1}:
   \begin{align}
&{\mathcal{W}_{conf}}[J_{\mathbb{O}},J_{\tilde{\mathbb{O}}},J_{\mathbb{S}},J_{\bar{\mathbb{S}}}]\nonumber\\
&=\frac{N^2-1}{2}\sum_{n=1}^{\infty}\frac{(-1)^{n}}{n}\Tr\Bigg(\frac{1}{2}i \square^{-1}(J_{\mathbb{O}_{s}}\otimes\mathcal{Y}_{s-2}^{\frac{5}{2}}+J_{\tilde{\mathbb{O}}_{s}}\otimes\mathcal{H}_{s-2}^{\frac{5}{2}})\Bigg)^n\nonumber\\
&\,+\frac{N^2-1}{2}\sum_{n=1}^{\infty}\frac{(-1)^{n}}{n}\Tr\Bigg(\frac{1}{2}i \square^{-1}(J_{\mathbb{O}_{s}}\otimes\mathcal{Y}_{s-2}^{\frac{5}{2}}-J_{\tilde{\mathbb{O}}_{s}}\otimes\mathcal{H}_{s-2}^{\frac{5}{2}})\Bigg)^n\nonumber\\
&\,+\frac{N^2-1}{2}\sum_{n=1}^{\infty}\frac{1}{n}\Tr\Bigg[\frac{1}{2}\left(I+\frac{1}{2}i \square^{-1}(J_{\mathbb{O}_{s}}\otimes\mathcal{Y}_{s-2}^{\frac{5}{2}}-J_{\tilde{\mathbb{O}}_{s}}\otimes\mathcal{H}_{s-2}^{\frac{5}{2}})\right)^{-1} i \square^{-1}J_{\bar{\mathbb{S}}_{s_1}} \otimes\mathcal{Y}_{s_1-2}^{\frac{5}{2}}\nonumber\\
&\quad\left(I+\frac{1}{2}i \square^{-1}(J_{\mathbb{O}_{s_2}}\otimes\mathcal{Y}_{s_2-2}^{\frac{5}{2}}+J_{\tilde{\mathbb{O}}_{s_2}}\otimes\mathcal{H}_{s_2-2}^{\frac{5}{2}}) \right)^{-1} i \square^{-1}J_{\mathbb{S}_{s_3}}\otimes\mathcal{Y}_{s_3-2}^{\frac{5}{2}} \Bigg]^n
\end{align}
Further expanding:
\begin{align}
	&\left(I+\frac{1}{2}i \square^{-1}(J_{\mathbb{O}_{s}}\otimes\mathcal{Y}_{s-2}^{\frac{5}{2}}\pm J_{\tilde{\mathbb{O}}_{s}}\otimes\mathcal{H}_{s-2}^{\frac{5}{2}}) \right)^{-1}\nonumber\\
	&= \sum_{n=0}^{\infty}(-1)^n \left(\frac{1}{2} i\square^{-1} (J_{\mathbb{O}_{s}}\otimes\mathcal{Y}_{s-2}^{\frac{5}{2}}\pm J_{\tilde{\mathbb{O}}_{s}}\otimes\mathcal{Y}_{s-2}^{\frac{5}{2}})\right)^n
\end{align}
we obtain:
   \begin{align}
	&{\mathcal{W}_{conf}}[J_{\mathbb{O}},J_{\tilde{\mathbb{O}}},J_{\mathbb{S}},J_{\bar{\mathbb{S}}}]\nonumber\\
	&=\frac{N^2-1}{2}\sum_{n=1}^{\infty}\frac{(-1)^{n}}{n}\Tr\Bigg(\frac{1}{2}i \square^{-1}(J_{\mathbb{O}_{s}}\otimes\mathcal{Y}_{s-2}^{\frac{5}{2}}+J_{\tilde{\mathbb{O}}_{s}}\otimes\mathcal{H}_{s-2}^{\frac{5}{2}})\Bigg)^n\nonumber\\
	&\,+\frac{N^2-1}{2}\sum_{n=1}^{\infty}\frac{(-1)^{n}}{n}\Tr\Bigg(\frac{1}{2}i \square^{-1}(J_{\mathbb{O}_{s}}\otimes\mathcal{Y}_{s-2}^{\frac{5}{2}}-J_{\tilde{\mathbb{O}}_{s}}\otimes\mathcal{H}_{s-2}^{\frac{5}{2}})\Bigg)^n\nonumber\\
	&\,+\frac{N^2-1}{2}\sum_{n=1}^{\infty}\frac{1}{n}\nonumber\\
	&\quad\Tr\Bigg[\frac{1}{2}\sum_{m_1=0}^{\infty}(-1)^{m_1} \left(\frac{1}{2} i\square^{-1} (J_{\mathbb{O}_{s}}\otimes\mathcal{Y}_{s-2}^{\frac{5}{2}}- J_{\tilde{\mathbb{O}}_{s}}\otimes\mathcal{Y}_{s-2}^{\frac{5}{2}})\right)^{m_1} i \square^{-1}J_{\bar{\mathbb{S}}_{s_1}} \otimes\mathcal{Y}_{s_1-2}^{\frac{5}{2}}\nonumber\\
	&\quad\sum_{m_2=0}^{\infty}(-1)^{m_2} \left(\frac{1}{2} i\square^{-1} (J_{\mathbb{O}_{s_2}}\otimes\mathcal{Y}_{s_2-2}^{\frac{5}{2}}+ J_{\tilde{\mathbb{O}}_{s_2}}\otimes\mathcal{Y}_{s_2-2}^{\frac{5}{2}})\right)^{m_2}i \square^{-1}J_{\mathbb{S}_{s_3}} \otimes\mathcal{Y}_{s_3-2}^{\frac{5}{2}} \Bigg]^n
\end{align}
Though the above expansion looks complicated, it suffices to observe that a generic term of the expansion has the structure:
\begin{align}
\Tr i\square^{-1} J \mathcal{Y}i\square^{-1}J \mathcal{Y}i\square^{-1}J \mathcal{Y}i\square^{-1}\ldots i\square^{-1} J \mathcal{H}i\square^{-1}J \mathcal{H}i\square^{-1}J \mathcal{H}i\square^{-1}\ldots i\square^{-1} J \mathcal{Y}i\square^{-1}J \mathcal{Y}
\end{align}
where the trace $\Tr$ includes all the objects to its right.
Therefore, the left derivatives in the products of $i\square^{-1}\mathcal{Y}i\square^{-1}$ or $i\square^{-1} \mathcal{H}i\square^{-1}$ may be rearranged thanks to the cyclicity of the trace to build the corresponding kernels of integral operators that only involve the right derivatives acting on the same $i\square^{-1}$, as in the previous example of the dictionary. \par
Analogously, after suitable a relabelling of the indices, the contribution of each trace to the generating functional may be rewritten in a form that matches the corresponding structure in terms of the kernel $\mathcal{D}^{-1}$ and the currents $j_{\mathcal{O}_{sk}}$ dual to the component operators $\mathcal{O}_{sk}$.

\subsection{Generating functional $\Gamma_{conf}$}

 Hence, the construction above may be condensed into the single formula:
   \begin{equation}
  \Gamma_{conf}[j_{\mathbb{O}},j_{\tilde{\mathbb{O}}},j_{\mathbb{S}},j_{\bar{\mathbb{S}}}] =-\frac{N^2-1}{2} \log\Det\boldsymbol{M}
  \end{equation} 
  with:
   \begin{align}
   & \boldsymbol{M}=\scalebox{1}{$
   	\begin{pmatrix}\mathbb{I}+\mathcal{D}^{-1}j_{\mathbb{O}}+\mathcal{D}^{-1}j_{\tilde{\mathbb{O}}}& \sqrt{2}\mathcal{D}^{-1}j_{\mathbb{S}}\\ \sqrt{2}\mathcal{D}^{-1}j_{\bar{\mathbb{S}}} &\mathbb{I}+\mathcal{D}^{-1}j_{\mathbb{O}}-\mathcal{D}^{-1}j_{\tilde{\mathbb{O}}} \end{pmatrix}$}
   \end{align}
 and:
 \begin{align}
	\label{kernelD_general}
	\mathcal{D}^{-1}_{s_1k_1,s_2k_2} &=\frac{i^{s_1}}{2}\frac{\Gamma(3)\Gamma(s_1+3)}{\Gamma(5)\Gamma(s_1+1)}{s_1\choose k_1}{s_2\choose k_2+2}(-\partial_{+})^{s_1-k_1+k_2}i\square^{-1}
\end{align}
Equivalently, we may employ the modified kernel in eq. \eqref{kernelD'}.
The above determinant reads (appendix \ref{A}):
    \begin{align}
  \label{genM}
  &\Gamma_{conf}[j_{\mathbb{O}},j_{\tilde{\mathbb{O}}},j_{\mathbb{S}},j_{\bar{\mathbb{S}}}] \nonumber\\
  &= 
  -\frac{N^2-1}{2}\Tr\log\left[\mathbb{I}+\mathcal{D}^{-1}j_{\mathbb{O}}+\mathcal{D}^{-1}j_{\tilde{\mathbb{O}}}\right]\nonumber\\
  &\quad-\frac{N^2-1}{2}\Tr\log\left[\mathbb{I}+\mathcal{D}^{-1}j_{\mathbb{O}}-\mathcal{D}^{-1}j_{\tilde{\mathbb{O}}}\right]\nonumber\\
   &\quad-\frac{N^2-1}{2}\Tr\log\left[\mathbb{I}-2\left(\mathbb{I}+\mathcal{D}^{-1}j_{\mathbb{O}}-\mathcal{D}^{-1}j_{\tilde{\mathbb{O}}}\right)^{-1}\mathcal{D}^{-1}j_{\bar{\mathbb{S}}}\left(\mathbb{I}+\mathcal{D}^{-1}j_{\mathbb{O}}+\mathcal{D}^{-1}j_{\tilde{\mathbb{O}}}\right)^{-1}\mathcal{D}^{-1}j_{\mathbb{S}}\right]
  \end{align}
The generating functional immediately reduces -- by setting the corresponding sources to zero -- to the generating functionals in the separate balanced and unbalanced sectors \cite{BPS1}:
      \begin{align}
 &\Gamma_{conf}[j_{\mathbb{O}},j_{\tilde{\mathbb{O}}}] = \Gamma_{conf}[j_{\mathbb{O}},j_{\tilde{\mathbb{O}}},0,0] \nonumber\\
 &=-\frac{N^2-1}{2}\Tr\log\left[\mathbb{I}+\mathcal{D}^{-1}j_{\mathbb{O}}-\mathcal{D}^{-1}j_{\tilde{\mathbb{O}}}\right]-\frac{N^2-1}{2}\Tr\log\left[\mathbb{I}+\mathcal{D}^{-1}j_{\mathbb{O}}+\mathcal{D}^{-1}j_{\tilde{\mathbb{O}}}\right]\nonumber\\
 &\Gamma_{conf}[j_{\mathbb{S}},j_{\bar{\mathbb{S}}}] =  \Gamma_{conf}[0,0,j_{\mathbb{S}},j_{\bar{\mathbb{S}}}] = -\frac{N^2-1}{2}\Tr\log\left[\mathbb{I}-2\mathcal{D}^{-1}j_{\bar{\mathbb{S}}}\mathcal{D}^{-1}j_{\mathbb{S}}\right]
  \end{align}
 Moreover, the generating functional of the Euclidean correlators reads:
    \begin{align}
	\label{genE}
	&\Gamma^E_{conf}[j_{\mathbb{O}^E},j_{\tilde{\mathbb{O}}^E},j_{\mathbb{S}^E},j_{\bar{\mathbb{S}}^E}] \nonumber\\
	&= 
	-\frac{N^2-1}{2}\Tr\log\left[\mathbb{I}+\mathcal{D}_E^{-1}j_{\mathbb{O}^E}+\mathcal{D}_E^{-1}j_{\tilde{\mathbb{O}}^E}\right]\nonumber\\
	&\quad-\frac{N^2-1}{2}\Tr\log\left[\mathbb{I}+\mathcal{D}_E^{-1}j_{\mathbb{O}^E}-\mathcal{D}_E^{-1}j_{\tilde{\mathbb{O}}^E}\right]\nonumber\\
	&\quad-\frac{N^2-1}{2}\Tr\log\Bigg[\mathbb{I}-2\left(\mathbb{I}+\mathcal{D}_E^{-1}j_{\mathbb{O}^E}-\mathcal{D}^{-1}j_{\tilde{\mathbb{O}^E}}\right)^{-1}\mathcal{D}_E^{-1}j_{\bar{\mathbb{S}}^E}\nonumber\\
	&\hspace{3.3cm}\left(\mathbb{I}+\mathcal{D}_E^{-1}j_{\mathbb{O}^E}+\mathcal{D}_E^{-1}j_{\tilde{\mathbb{O}}^E}\right)^{-1}\mathcal{D}_E^{-1}j_{\mathbb{S}^E}\Bigg]
\end{align}
with \cite{BPS1}:
\begin{align} \label{kerDE}
\nonumber\\
{\mathcal{D}^{-1}_E}_{s_1k_1,s_2k_2}&=\frac{(-i)^{k_1-k_2}}{2}\frac{\Gamma(3)\Gamma(s_1+3)}{\Gamma(5)\Gamma(s_1+1)}{s_1\choose k_1}{s_2\choose k_2+2}\partial_{z}^{s_1-k_1+k_2}\Laplace^{-1}
\end{align}
obtained by Wick rotation \cite{BPS1}:
\begin{align} 
  \label{Wick}
&x^+= \frac{x^0 + x^3}{\sqrt{2}}
\rightarrow -ix^z= -i \frac{x^4+ix^3}{\sqrt{2}}
\end{align}
where:
\begin{equation}
	\Laplace= \delta_{\mu\nu}\partial_{\mu}\partial_{\nu}=\partial_4^2+\sum_{i=1}^{3}\partial_i^2
\end{equation}
and \cite{BPS1}:
\begin{equation} \label{Laplacian}
\Laplace^{-1}(x-y)=-	\frac{1}{4\pi^2}\frac{1}{(x-y)^2}
\end{equation}
Equivalently, we may employ the modified kernel:
\begin{align} 
{\mathcal{D'}^{-1}_E}_{s_1k_1,s_2k_2}&=\frac{1}{2}\frac{\Gamma(3)\Gamma(s_1+3)}{\Gamma(5)\Gamma(s_1+1)}{s_1\choose k_1}{s_2\choose k_2+2}\partial_{z}^{s_1-k_1+k_2}\Laplace^{-1}
\end{align}

\section{Generating functional in the momentum representation\label{momgen}}

The generating functional in the momentum representation is defined by the functional integral:
  \begin{align}
	\mathcal{Z}_{conf}[J_{\mathcal{O}}] =& \int \mathcal{D}A\mathcal{D}\bar{A}\, e^{-i \int d^4x\, \bar{A}^a \square A^a}\exp\Bigg(\int \frac{d^4p}{(2\pi)^4}\sum_i\, J_{\mathcal{O}_i}(-p)\mathcal{O}_i(p)\Bigg)
\end{align}
Correspondingly, the correlators read:
\begin{align}
\langle \mathcal{O}_{s_1}(p_1)\ldots\mathcal{O}_{s_n}(p_n)\rangle = (2\pi)^4\frac{\delta }{\delta J_{\mathcal{O}_{s_1}}(-p_1)}\cdots (2\pi)^4\frac{\delta }{\delta J_{\mathcal{O}_{s_n}}(-p_n)}\mathcal{W}_{conf}[J_{\mathcal{O}}] 
\end{align} 
Employing the dictionary, we get as well:
\begin{align}
	\label{momentumcorr}
	&\langle \mathcal{O}_{s_1}(p_1)\ldots\mathcal{O}_{s_n}(p_n)\rangle \nonumber\\
	&= \sum_{k_1=0}^{l_1}(2\pi)^4\frac{\delta }{\delta j_{\mathcal{O}_{s_1k_1}}(-p_1)}\cdots \sum_{k_n=0}^{l_n}(2\pi)^4\frac{\delta }{\delta j_{\mathcal{O}_{s_nk_n}}(-p_n)}\Gamma_{conf}[j_{\mathcal{O}}] 
\end{align} 

\subsection{Standard basis}

To find the explicit form of the generating functional in the momentum representation we choose again as example:
\begin{equation}
	 \Gamma_{conf}[j_{\mathbb{O}}] =-(N^2-1) \log \Det \left(\delta_{s_1k_1,s_2k_2}\delta^{(4)}(x-y)+\mathcal{D}^{-1}_{s_1k_1,s_2k_2}(x-y)j_{\mathbb{O}_{s_2k_2}}(y)\right)
\end{equation}
where the argument of the determinant is the kernel:
\begin{equation} \label{KK}
	K_{s_1k_1,s_2k_2}(x,y)=\delta_{s_1k_1,s_2k_2}\delta^{(4)}(x-y)+\mathcal{D}^{-1}_{s_1k_1,s_2k_2}(x-y)j_{\mathbb{O}_{s_2k_2}}(y)
\end{equation}
of the integral operator:
\begin{equation}
	\psi_{s_1k_1}(x) =\sum_{s_2k_2} \int K_{s_1k_1,s_2k_2}(x,y) \phi_{s_2k_2}(y) d^4y
\end{equation}
In order to obtain the kernel in the momentum representation \cite{BPS1} we perform the Fourier transform of the lhs:
\begin{equation}
	\psi_{s_1k_1}(q)=\int  \psi_{s_1k_1}(x) e^{-iq\cdot x} d^4x=\sum_{s_2k_2} \int K_{s_1k_1,s_2k_2}(x,y) \phi_{s_2k_2}(y) e^{-iq\cdot x}\,d^4x\,d^4y
\end{equation}
and write the rhs in terms of the Fourier-transformed fields:
\begin{equation}
	\psi_{s_1k_1}(q) =\sum_{s_2k_2} \int K_{s_1k_1,s_2k_2}(x,y) \phi_{s_2k_2}(p)e^{ip\cdot y} e^{-iq\cdot x}\,d^4x\,d^4y\frac{d^4p}{(2\pi)^4}
\end{equation}
Substituting the kernel in eq. \eqref{KK}:
\begin{align}
	\psi_{s_1k_1}(q) =&\sum_{s_2k_2} \int \delta_{s_1k_1,s_2k_2}\delta^{(4)}(x-y)\phi_{s_2k_2}(p)e^{ipy} e^{-iqx}\nonumber\\
	&+ \mathcal{D}^{-1}_{s_1k_1,s_2k_2}(x-y)j_{\mathbb{O}_{s_2k_2}}(y) \phi_{s_2k_2}(p)e^{ipy} e^{-iqx}\,d^4x\,d^4y\frac{d^4p}{(2\pi)^4}
\end{align}
we get:
\begin{align}
	\psi_{s_1k_1}(q) =&\sum_{s_2k_2} \int \delta_{s_1k_1,s_2k_2}(2\pi)^4\delta^{(4)}(p-q)\phi_{s_2k_2}(p)\frac{d^4p}{(2\pi)^4}\nonumber\\
	&+ \sum_{s_2k_2} \int \mathcal{D}^{-1}_{s_1k_1,s_2k_2}(x-y)j_{\mathbb{O}_{s_2k_2}}(y) \phi_{s_2k_2}(p)e^{ip\cdot y} e^{-iq\cdot x}\,d^4x\,d^4y\frac{d^4p}{(2\pi)^4}
\end{align}
The second line in the above equation becomes:
\begin{align}
	\sum_{s_2k_2} \int\mathcal{D}^{-1}_{s_1k_1,s_2k_2}(k_1)j_{\mathbb{O}_{s_2k_2}}(k_2) \phi_{s_2k_2}(p)e^{ik_1\cdot (x-y)}e^{ik_2\cdot y}e^{ip\cdot y} e^{-iq\cdot x}\,d^4x\,d^4y\frac{d^4p}{(2\pi)^4}\frac{d^4k_1}{(2\pi)^4}\frac{d^4k_2}{(2\pi)^4}
\end{align}
that reduces to:
\begin{align}
	\sum_{s_2k_2} \int\mathcal{D}^{-1}_{s_1k_1,s_2k_2}(q)j_{\mathbb{O}_{s_2k_2}}(q-p) \phi_{s_2k_2}(p)\frac{d^4p}{(2\pi)^4}
\end{align}
Therefore, the kernel in the momentum representation is:
\begin{equation}
	K_{s_1k_1,s_2k_2}(q_1,q_2)=\delta_{s_1k_1,s_2k_2}(2\pi)^4\delta^{(4)}(q_1-q_2)+\mathcal{D}^{-1}_{s_1k_1,s_2k_2}(q_1)j_{\mathbb{O}_{s_2k_2}}(q_1-q_2)
\end{equation}
that defines the integral operator:
\begin{equation}
	\psi_{s_1k_1}(q_1) =\sum_{s_2k_2} \int K_{s_1k_1,s_2k_2}(q_1,q_2) \phi_{s_2k_2}(q_2) \frac{d^4q_2}{(2\pi)^4}
\end{equation}
with:
\begin{align}
	\mathcal{D}^{-1}_{s_1k_1,s_2k_2}(p) &=\frac{i^{s_1}}{2}\frac{\Gamma(3)\Gamma(s_1+3)}{\Gamma(5)\Gamma(s_1+1)}{s_1\choose k_1}{s_2\choose k_2+2}(-ip_{+})^{s_1-k_1+k_2}\frac{-i}{\rvert p\rvert^2+i\epsilon} 
\end{align}
Equivalently, we may employ the kernel:
\begin{align}
	\mathcal{D'}^{-1}_{s_1k_1,s_2k_2}(p) &=\frac{1}{2}\frac{\Gamma(3)\Gamma(s_1+3)}{\Gamma(5)\Gamma(s_1+1)}{s_1\choose k_1}{s_2\choose k_2+2}p_{+}^{s_1-k_1+k_2}\frac{-i}{\rvert p\rvert^2+i\epsilon} 
\end{align}
Hence, the generating functional in the momentum representation reads \cite{BPS1}:
\begin{equation}
	\Gamma_{conf}[j_{\mathbb{O}}] = -(N^2-1)  \log\Det\left(\delta_{s_1k_1,s_2k_2} (2\pi)^4\delta^{(4)}(q_1-q_2)+\mathcal{D}^{-1}_{s_1k_1,s_2k_2}(q_1)j_{\mathbb{O}_{s_2k_2}}(q_1-q_2)\right)
\end{equation}
In order to obtain the correlators we expand the generating functional:
\begin{align} \label{gen}
	\nonumber
	\Gamma_{conf}[j_{\mathbb{O}}]  &=-(N^2-1) \sum_{n=1}^{\infty}\frac{(-1)^{n+1}}{n}\int\frac{d^4q_1}{(2\pi)^4}\cdots \frac{d^4q_n}{(2\pi)^4}\sum_{s_1k_1}\ldots\sum_{s_nk_n}\\
	&\quad\mathcal{D}^{-1}_{s_1k_1,s_2k_2}(q_1)j_{\mathbb{O}_{s_2k_2}}(q_1-q_2)\mathcal{D}^{-1}_{s_2k_2,s_3k_3}(q_2)j_{\mathbb{O}_{s_3k_3}}(q_2-q_3)\nonumber\\
	&\quad\ldots \mathcal{D}^{-1}_{s_nk_n,s_1k_1}(q_n)j_{\mathbb{O}_{s_1k_1}}(q_n-q_1)
\end{align}
We change variables defining $p_i = q_i-q_{i+1}$  with $q_{n+1} \equiv q_1$ that automatically ensures $\sum_{i = 1}^{n} p_i = 0$:
 \begin{align}
 &q_1 = q_n-p_n\nonumber\\
 &q_2=q_1-p_1 = q_n-p_n-p_1\nonumber\\
&q_3 = q_2-p_2 = q_n-p_n-p_2-p_1\nonumber\\
&\ldots\nonumber\\
&q_{n-1} =q_n-p_n-\sum_{i = 1}^{n-2} p_i\nonumber\\
&q_n = q_{n-1}-p_{n-1}
\end{align}
so that, setting $q_n\equiv q$ and inserting in the rhs of eq. \eqref{gen} the integral $\int \frac{d^4p_n}{(2\pi)^4}\,(2\pi)^4\delta^{(4)}(p_1+p_2+\cdots+p_n)$ to keep $p_n$ while enforcing $\sum_{i = 1}^{n} p_i = 0$, we obtain:
\begin{align}
	\nonumber
	\Gamma_{conf}[j_{\mathbb{O}}]  =&-(N^2-1) \sum_{n=1}^{\infty}\frac{(-1)^{n+1}}{n}\int\frac{d^4p_1}{(2\pi)^4}\cdots \frac{d^4p_n}{(2\pi)^4}\,(2\pi)^4\delta^{(4)}(p_1+p_2+\cdots+p_n)\\\nonumber
	&\int\frac{d^4q}{(2\pi)^4} \sum_{s_1k_1}\ldots\sum_{s_nk_n}\mathcal{D}^{-1}_{s_1k_1,s_2k_2}(q-p_n)j_{\mathbb{O}_{s_2k_2}}(p_1)\\
	&\mathcal{D}^{-1}_{s_2k_2,s_3k_3}(q-p_1-p_n)j_{\mathbb{O}_{s_3k_3}}(p_2)\ldots \mathcal{D}^{-1}_{s_nk_n,s_1k_1}(q)j_{\mathbb{O}_{s_1k_1}}(p_n)
\end{align}
Moreover, making the substitution $p_i\rightarrow-p_i$, we get:
\begin{align}
	\nonumber
	\Gamma_{conf}[j_{\mathbb{O}}]  =&-(N^2-1) \sum_{n=1}^{\infty}\frac{(-1)^{n+1}}{n}\int\frac{d^4p_1}{(2\pi)^4}\cdots \frac{d^4p_n}{(2\pi)^4}\,(2\pi)^4\delta^{(4)}(p_1+p_2+\cdots+p_n)\\\nonumber
	&\int\frac{d^4q}{(2\pi)^4} \sum_{s_1k_1}\ldots\sum_{s_nk_n}\mathcal{D}^{-1}_{s_1k_1,s_2k_2}(q+p_n)j_{\mathbb{O}_{s_2k_2}}(-p_1)\\
	&\mathcal{D}^{-1}_{s_2k_2,s_3k_3}(q+p_1+p_n)j_{\mathbb{O}_{s_3k_3}}(-p_2)\ldots \mathcal{D}^{-1}_{s_nk_n,s_1k_1}(q)j_{\mathbb{O}_{s_1k_1}}(-p_n)
\end{align}
We employ eq. (\ref{momentumcorr}) to obtain the correlators:
\begin{align}
	\langle \mathbb{O}_{s_1}(p_1)\ldots\mathbb{O}_{s_n}(p_n)\rangle &=\frac{N^2-1}{2^n}\frac{(-1)^{n}}{n}\,(2\pi)^4\delta^{(4)}(p_1+p_2+\cdots+p_n)\nonumber\\
	&\quad\sum_{\sigma \in P_n}\int\frac{d^4q}{(2\pi)^4} \,\mathcal{D}^{-1}_{s_{\sigma(1)}k_{\sigma(1)},s_{\sigma(2)}k_{\sigma(2)}}(q+p_{\sigma(n)})\nonumber\\
	&\quad\mathcal{D}^{-1}_{s_{\sigma(2)}k_{\sigma(2)},s_{\sigma(3)}k_{\sigma(3)}}(q+p_{\sigma(1)}+p_{\sigma(n)})\ldots \mathcal{D}^{-1}_{s_{\sigma(n)}k_{\sigma(n)},s_{\sigma(1)}k_{\sigma(1)}}(q)
\end{align} 
Explicitly, by eq. (\ref{kernelD}):
\begin{align}
	\nonumber
	&\hspace{-0.15cm}\left(\frac{2\Gamma(5)\Gamma(s_1+1)}{\Gamma(3)\Gamma(s_1+3)}\right)\ldots\left( \frac{2\Gamma(5)\Gamma(s_{n}+1)}{\Gamma(3)\Gamma(s_{n}+3)}\right)\\\nonumber
	&\langle \mathbb{O}_{s_1}(p_1)\ldots \mathbb{O}_{s_n}(p_n)\rangle \\\nonumber
	&=(N^2-1)(2\pi)^4 i^{n}\delta^{(4)}(p_1+\cdots+p_{n})\\\nonumber
	&\quad\sum_{k_1=0}^{s_1-2}\ldots \sum_{k_{n} = 0}^{s_{n}-2}{s_1\choose k_1}{s_1\choose k_1+2}\ldots {s_{n}\choose k_{n}}{s_{n}\choose k_{n}+2}\\\nonumber
	&\quad\frac{1}{n}\sum_{\sigma\in P_{n}}\int \frac{d^4 q}{(2\pi)^4}\frac{(p_{\sigma(1)}+q)_{+}^{s_{\sigma(1)}-k_{\sigma(1)}+k_{\sigma(2)}}}{\rvert p_{\sigma(1)}+q\rvert^2}\frac{(p_{\sigma(1)}+p_{\sigma(2)}+q)_{+}^{s_{\sigma(2)}-k_{\sigma(2)}+k_{\sigma(3)}}}{\rvert p_{\sigma(1)}+p_{\sigma(2)}+q\rvert^2}\\
	&\quad\cdots\frac{(\sum_{l=1}^{n-1} p_{\sigma(l)}+q)_{+}^{s_{\sigma(n-1)}-k_{\sigma(n-1)}+k_{\sigma(n)}}}{\rvert\sum_{l=1}^{n-1} p_{\sigma(l)}+q\rvert^2}\frac{(q)_{+}^{s_{\sigma(n)}-k_{\sigma(n)}+k_{\sigma(1)}}}{\rvert q\rvert^2}
\end{align}
that coincides with the computation in \cite{BPS1}. More generally, the generating functional in the momentum representation reads from eqs. \eqref{wexplicit} and \eqref{genM} with (minus) the propagator in the momentum representation:
\begin{equation} \label{AB}
	i\square^{-1}(p)=\frac{-i}{\rvert p\rvert^2+i\epsilon} 
\end{equation}
and correspondingly:
 \begin{equation} \label{BA}
 (i\partial_+)^{s-k_1+k_2} i\square^{-1}(p)= (-p_+)^{s-k_1+k_2} i\square^{-1}(p)
\end{equation}

\section{Mixed $3$- and $4$-point conformal correlators \label{s5}}

We employ the generating functional to calculate the $3$- and $4$-point correlators in the mixed balanced/unbalanced sector.

\subsection{$3$-point correlators in Minkowskian space-time}

 The generating functional of the mixed $\mathbb{O}$, $\mathbb{S}$ and $\bar{\mathbb{S}}$ correlators reads:
    \begin{align}
	&\Gamma_{conf}[j_{\mathbb{O}},0,j_{\mathbb{S}},j_{\bar{\mathbb{S}}}]=-\frac{N^2-1}{2}\Tr\log\left[\mathbb{I}-2(\mathbb{I}+\mathcal{D}^{-1}j_{\mathbb{O}})^{-1}\mathcal{D}^{-1}j_{\bar{\mathbb{S}}}\left(\mathbb{I}+\mathcal{D}^{-1}j_{\mathbb{O}}\right)^{-1}\mathcal{D}^{-1}j_{\mathbb{S}}\right]
\end{align}
Only two terms contribute to the mixed $3$-point correlator $\langle\mathbb{O}_{s_1}(x)\mathbb{S}_{s_2}(y)\bar{\mathbb{S}}_{s_3}(z)\rangle$:
\begin{align}
	&\Gamma_{conf}[j_{\mathbb{O}},0,j_{\mathbb{S}},j_{\bar{\mathbb{S}}}]\nonumber\\
	&=\frac{N^2-1}{2}2\Tr\left((\mathbb{I}+\mathcal{D}^{-1}j_{\mathbb{O}})^{-1}\mathcal{D}^{-1}j_{\bar{\mathbb{S}}}\left(\mathbb{I}+\mathcal{D}^{-1}j_{\mathbb{O}}\right)^{-1}\mathcal{D}^{-1}j_{\mathbb{S}}\right)+\cdots\nonumber\\
	&=\frac{N^2-1}{2}2\Tr\left((\mathbb{I}-\mathcal{D}^{-1}j_{\mathbb{O}})\mathcal{D}^{-1}j_{\bar{\mathbb{S}}}\left(\mathbb{I}-\mathcal{D}^{-1}j_{\mathbb{O}}\right)\mathcal{D}^{-1}j_{\mathbb{S}}\right)+\cdots\nonumber\\
	&=-\frac{N^2-1}{2}2\Tr\left(\mathcal{D}^{-1}j_{\mathbb{O}}\mathcal{D}^{-1}j_{\bar{\mathbb{S}}}\mathcal{D}^{-1}j_{\mathbb{S}}\right)-\frac{N^2-1}{2}2\Tr\left(\mathcal{D}^{-1}j_{\bar{\mathbb{S}}}\mathcal{D}^{-1}j_{\mathbb{O}}\mathcal{D}^{-1}j_{\mathbb{S}}\right)+\cdots
\end{align}
 Hence:
\begin{align}
&\langle\mathbb{O}_{s_1}(x)\mathbb{S}_{s_2}(y)\bar{\mathbb{S}}_{s_3}(z)\rangle \nonumber\\
&= \sum_{k_1=0}^{s_1-2}\sum_{k_2=0}^{s_2-2}\sum_{k_3=0}^{s_3-2}\frac{\delta}{\delta j_{\mathbb{O}_{s_1k_1}}(x)}\frac{\delta}{\delta j_{\mathbb{S}_{s_2k_2}}(y)}\frac{\delta}{\delta j_{\bar{\mathbb{S}}_{s_3k_3}}(z)} \Gamma_{conf}[j_{\mathbb{O}},j_{\tilde{\mathbb{O}}},j_{\mathbb{S}},j_{\bar{\mathbb{S}}}]\nonumber\\
&=-\frac{N^2-1}{2}2\sum_{k_1=0}^{s_1-2}\sum_{k_2=0}^{s_2-2}\sum_{k_3=0}^{s_3-2}\frac{\delta}{\delta j_{\mathbb{O}_{s_1k_1}}(x)}\frac{\delta}{\delta j_{\mathbb{S}_{s_2k_2}}(y)}\frac{\delta}{\delta j_{\bar{\mathbb{S}}_{s_3k_3}}(z)}\nonumber\\
&\quad\int d^4x_1d^4x_2d^4x_3\sum_{k'_1s'_1}\sum_{k'_2s'_2}\sum_{k'_3s'_3}\Bigg(\mathcal{D}^{-1}_{s'_1k'_1,s'_2k'_2}(x_1-x_2) j_{\mathbb{O}_{s'_2k'_2}}(x_2)\nonumber\\
&\quad\mathcal{D}^{-1}_{s'_2k'_2,s'_3k'_3}(x_2-x_3) j_{\mathbb{S}_{s'_3k'_3}}(x_3)\mathcal{D}^{-1}_{s'_3k'_3,s'_1k'_1}(x_3-x_1) j_{\bar{\mathbb{S}}_{s'_1k'_1}}(x_1)\nonumber\\
&\quad+\mathcal{D}^{-1}_{s'_1k'_1,s'_2k'_2}(x_1-x_2) j_{\bar{\mathbb{S}}_{s'_2k'_2}}(x_2)\mathcal{D}^{-1}_{s'_2k'_2,s'_3k'_3}(x_2-x_3) j_{\mathbb{O}_{s'_3k'_3}}(x_3)\nonumber\\
&\quad\mathcal{D}^{-1}_{s'_3k'_3,s'_1k'_1}(x_3-x_1) j_{{\mathbb{S}}_{s'_1k'_1}}(x_1)\Bigg)\nonumber\\
&=-(N^2-1)\sum_{k_1=0}^{s_1-2}\sum_{k_2=0}^{s_2-2}\sum_{k_3=0}^{s_3-2}\Big(\mathcal{D}^{-1}_{s_1k_1,s_2k_2}(x-y)\mathcal{D}^{-1}_{s_2k_2,s_3k_3}(y-z)\mathcal{D}^{-1}_{s_3k_3,s_1k_1}(z-x)\nonumber\\
&\quad+\mathcal{D}^{-1}_{s_2k_2,s_3k_3}(y-z)\mathcal{D}^{-1}_{s_3k_3,s_1k_1}(z-x)\mathcal{D}^{-1}_{s_1k_1,s_2k_2}(x-y)\Big)\nonumber\\
&=-2(N^2-1)\sum_{k_1=0}^{s_1-2}\sum_{k_2=0}^{s_2-2}\sum_{k_3=0}^{s_3-2}\mathcal{D}^{-1}_{s_1k_1,s_2k_2}(x-y)\mathcal{D}^{-1}_{s_2k_2,s_3k_3}(y-z)\mathcal{D}^{-1}_{s_3k_3,s_1k_1}(z-x)
\end{align}
Employing eq. (\ref{kernelD}), we obtain:
\begin{align}
\nonumber
\langle\mathbb{O}_{s_1}(x)\mathbb{S}_{s_2}(y)\bar{\mathbb{S}}_{s_3}(z)\rangle &=-\frac{1}{(4\pi^2)^3} 2 \left(\frac{2}{4!}\right)^3\frac{N^2-1}{8}i^{s_1+s_2+s_3}2^{s_1+s_2+s_3}\\\nonumber
&\quad(s_1+1)(s_1+2)(s_2+1)(s_2+2)(s_3+1)(s_3+2)\\\nonumber
&\quad\sum_{k_1 = 0}^{s_1-2}\sum_{k_2 = 0}^{s_2-2}\sum_{k_3 = 0}^{s_3-2}{s_1\choose k_1}{s_1\choose k_1+2}{s_2\choose k_2}{s_2\choose k_2+2}{s_3\choose k_3}{s_3\choose k_3+2}\\\nonumber
&\quad(s_1-k_1+k_2)!(s_2-k_2+k_3)!(s_3-k_3+k_1)!  \\
&\quad\frac{(x-y)^{s_1-k_1+k_2}_+}{(\rvert x-y\rvert^2)^{s_1+1-k_1+k_2}}\frac{(y-z)^{s_2-k_2+k_3}_+}{(\rvert y-z\rvert^2)^{s_2+1-k_2+k_3}}\frac{(z-x)^{s_3-k_3+k_1}_+}{(\rvert z-x\rvert^2)^{s_3+1-k_3+k_1}}
\end{align}
that agrees with \cite{BPS1}. We also infer from the generating functional that \cite{BPS1}:
\begin{equation}
\langle \tilde{\mathbb{O}}_{s_1}(x)\mathbb{S}_{s_2}(y)\bar{\mathbb{S}}_{s_3}(z)\rangle = 0
\end{equation}
Indeed:
    \begin{align}
	&\Gamma_{conf}[0,j_{\tilde{\mathbb{O}}},j_{\mathbb{S}},j_{\bar{\mathbb{S}}}]=-\frac{N^2-1}{2}\Tr\log\left[\mathbb{I}-2(\mathbb{I}-\mathcal{D}^{-1}j_{\tilde{\mathbb{O}}})^{-1}\mathcal{D}^{-1}j_{\bar{\mathbb{S}}}\left(\mathbb{I}+\mathcal{D}^{-1}j_{\tilde{\mathbb{O}}}\right)^{-1}\mathcal{D}^{-1}j_{\mathbb{S}}\right]
\end{align}
contributes two terms linear in $j_{\tilde{\mathbb{O}}}$ with opposite sign that cancel each other.

\subsection{$4$-point correlators in Minkowskian space-time}

Only three terms contribute to the mixed $4$-point correlator $\langle\mathbb{O}_{s_1}(x_1)\mathbb{O}_{s_2}(x_2)\mathbb{S}_{s_3}(x_3)\bar{\mathbb{S}}_{s_4}(x_4)\rangle$:
    \begin{align}
	&\Gamma_{conf}[j_{\mathbb{O}},0,j_{\mathbb{S}},j_{\bar{\mathbb{S}}}]\nonumber\\
	&=\frac{N^2-1}{2}2\Tr\left((\mathbb{I}-\mathcal{D}^{-1}j_{\mathbb{O}}+\mathcal{D}^{-1}j_{\mathbb{O}}\mathcal{D}^{-1}j_{\mathbb{O}})\mathcal{D}^{-1}j_{\bar{\mathbb{S}}}\left(\mathbb{I}-\mathcal{D}^{-1}j_{\mathbb{O}}+\mathcal{D}^{-1}j_{\mathbb{O}}\mathcal{D}^{-1}j_{\mathbb{O}}\right)\mathcal{D}^{-1}j_{\mathbb{S}}\right)+\cdots\nonumber\\
	&=\frac{N^2-1}{2}2\Tr\left(\mathcal{D}^{-1}j_{\mathbb{O}}\mathcal{D}^{-1}j_{\bar{\mathbb{S}}}\mathcal{D}^{-1}j_{\mathbb{O}}\mathcal{D}^{-1}j_{\mathbb{S}}\right)+\frac{N^2-1}{2}2\Tr\left(\mathcal{D}^{-1}j_{\mathbb{O}}\mathcal{D}^{-1}j_{\mathbb{O}}\mathcal{D}^{-1}j_{\bar{\mathbb{S}}}\mathcal{D}^{-1}j_{\mathbb{S}}\right)\nonumber\\
	&\quad+\frac{N^2-1}{2}2\Tr\left(\mathcal{D}^{-1}j_{\bar{\mathbb{S}}}\mathcal{D}^{-1}j_{\mathbb{O}}\mathcal{D}^{-1}j_{\mathbb{O}}\mathcal{D}^{-1}j_{\mathbb{S}}\right)+\cdots
\end{align}
Hence:
\begin{align}
&\langle\mathbb{O}_{s_1}(x_1)\mathbb{O}_{s_2}(x_2)\mathbb{S}_{s_3}(x_3)\bar{\mathbb{S}}_{s_4}(x_4)\rangle\nonumber\\
&= \sum_{k_1=0}^{s_1-2}\sum_{k_2=0}^{s_2-2}\sum_{k_3=0}^{s_3-2}\sum_{k_4=0}^{s_4-2}\frac{\delta}{\delta j_{\mathbb{O}_{s_1k_1}}(x_1)}\frac{\delta}{\delta j_{\mathbb{O}_{s_2k_2}}(x_2)}\frac{\delta}{\delta j_{\mathbb{S}_{s_3k_3}}(x_3)}\frac{\delta}{\delta j_{\bar{\mathbb{S}}_{s_4k_4}}(x_4)} \Gamma_{conf}[j_{\mathbb{O}},j_{\tilde{\mathbb{O}}},j_{\mathbb{S}},j_{\bar{\mathbb{S}}}]\nonumber\\
&=\frac{N^2-1}{2} 2\sum_{k_1=0}^{s_1-2}\sum_{k_2=0}^{s_2-2}\sum_{k_3=0}^{s_3-2}\sum_{k_4=0}^{s_4-2}\frac{\delta}{\delta j_{\mathbb{O}_{s_1k_1}}(x_1)}\frac{\delta}{\delta j_{\mathbb{O}_{s_2k_2}}(x_2)}\frac{\delta}{\delta j_{\mathbb{S}_{s_3k_3}}(x_3)}\frac{\delta}{\delta j_{\bar{\mathbb{S}}_{s_4k_4}}(x_4)}\nonumber\\
&\quad\int d^4y_1d^4y_2d^4y_3d^4y_4\sum_{k'_1s'_1}\sum_{k'_2s'_2}\sum_{k'_3s'_3}\sum_{k'_4s'_4}\nonumber\\
&\quad\Bigg(\mathcal{D}^{-1}_{s'_1k'_1,s'_2k'_2}(y_1-y_2) j_{\mathbb{O}_{s'_2k'_2}}(y_2)\mathcal{D}^{-1}_{s'_2k'_2,s'_3k'_3}(y_2-y_3) j_{\bar{\mathbb{S}}_{s'_3k'_3}}(y_3)\nonumber\\
&\quad\quad\mathcal{D}^{-1}_{s'_3k'_3,s'_4k'_4}(y_3-y_4) j_{\mathbb{O}_{s'_4k'_4}}(y_4)\mathcal{D}^{-1}_{s'_4k'_4,s'_1k'_1}(y_4-y_1) j_{\mathbb{S}_{s'_1k'_1}}(y_1)\nonumber\\
&\quad+\mathcal{D}^{-1}_{s'_1k'_1,s'_2k'_2}(y_1-y_2) j_{\mathbb{O}_{s'_2k'_2}}(y_2)\mathcal{D}^{-1}_{s'_2k'_2,s'_3k'_3}(y_2-y_3) j_{\mathbb{O}_{s'_3k'_3}}(y_3)\nonumber\\
&\quad\quad\mathcal{D}^{-1}_{s'_3k'_3,s'_4k'_4}(y_3-y_4) j_{\bar{\mathbb{S}}_{s'_4k'_4}}(y_4)\mathcal{D}^{-1}_{s'_4k'_4,s'_1k'_1}(y_4-y_1) j_{\mathbb{S}_{s'_1k'_1}}(y_1)\nonumber\\
&\quad+\mathcal{D}^{-1}_{s'_1k'_1,s'_2k'_2}(y_1-y_2) j_{\mathbb{O}_{s'_2k'_2}}(y_2)\mathcal{D}^{-1}_{s'_2k'_2,s'_3k'_3}(y_2-y_3) j_{\mathbb{O}_{s'_3k'_3}}(y_3)\nonumber\\
&\quad\quad\mathcal{D}^{-1}_{s'_3k'_3,s'_4k'_4}(y_3-y_4) j_{\mathbb{S}_{s'_4k'_4}}(y_4)\mathcal{D}^{-1}_{s'_4k'_4,s'_1k'_1}(y_4-y_1) j_{\bar{\mathbb{S}}_{s'_1k'_1}}(y_1)\Bigg)
\end{align}
Performing the functional derivatives and employing the symmetry properties of the Gegenbauer polynomials (appendix \ref{B}), we get:
\begin{align}
	&\langle\mathbb{O}_{s_1}(x_1)\mathbb{O}_{s_2}(x_2)\mathbb{S}_{s_3}(x_3)\bar{\mathbb{S}}_{s_4}(x_4)\rangle\nonumber\\
	&= \sum_{k_1=0}^{s_1-2}\sum_{k_2=0}^{s_2-2}\sum_{k_3=0}^{s_3-2}\sum_{k_4=0}^{s_4-2}\frac{\delta}{\delta j_{\mathbb{O}_{s_1k_1}}(x_1)}\frac{\delta}{\delta j_{\mathbb{O}_{s_2k_2}}(x_2)}\frac{\delta}{\delta j_{\mathbb{S}_{s_3k_3}}(x_3)}\frac{\delta}{\delta j_{\bar{\mathbb{S}}_{s_4k_4}}(x_4)} \Gamma_{conf}[j_{\mathbb{O}},j_{\tilde{\mathbb{O}}},j_{\mathbb{S}},j_{\bar{\mathbb{S}}}]\nonumber\\
	&=(N^2-1)\sum_{k_1=0}^{s_1-2}\sum_{k_2=0}^{s_2-2}\sum_{k_3=0}^{s_3-2}\sum_{k_4=0}^{s_4-2}\sum_{\sigma\in P_2}\nonumber\\
	&\quad\Bigg(\mathcal{D}^{-1}_{s_3k_3,s_{\sigma(1)}k_{\sigma(1)}}(x_3-x_{\sigma(1)}) \mathcal{D}^{-1}_{s_{\sigma(1)}k_{\sigma(1)},s_4k_4}(x_{\sigma(1)}-x_4)\nonumber\\
	&\quad\quad\mathcal{D}^{-1}_{s_4k_4,s_{\sigma(2)}k_{\sigma(2)}}(x_4-x_{\sigma(2)}) \mathcal{D}^{-1}_{s_{\sigma(2)}k_{\sigma(2)},s_3k_3}(x_{\sigma(2)}-x_3) \nonumber\\
	&\quad+2\mathcal{D}^{-1}_{s_3k_3,s_{\sigma(1)}k_{\sigma(1)}}(x_3-x_{\sigma(1)}) \mathcal{D}^{-1}_{s_{\sigma(1)}k_{\sigma(1)},s_{\sigma(2)}k_{\sigma(2)}}(x_{\sigma(1)}-x_{\sigma(2)}) \nonumber\\
	&\quad\quad\mathcal{D}^{-1}_{s_{\sigma(2)}k_{\sigma(2)},s_4k_4}(x_{\sigma(2)}-x_4) \mathcal{D}^{-1}_{s_4k_4,s_3k_3}(x_4-x_3)
	 \Bigg)
\end{align} 
Explicitly:
\begin{align}
	&2^4\frac{\Gamma(5)\Gamma(s_1+1)}{\Gamma(3)\Gamma(s_1+3)}
	\frac{\Gamma(5)\Gamma(s_2+1)}{\Gamma(3)\Gamma(s_2+3)}\frac{\Gamma(5)\Gamma(s_3+1)}{\Gamma(3)\Gamma(s_3+3)}\frac{\Gamma(5)\Gamma(s_4+1)}{\Gamma(3)\Gamma(s_4+3)}\nonumber\\
	&\langle\mathbb{O}_{s_1}(x_1)\mathbb{O}_{s_2}(x_2)\mathbb{S}_{s_3}(x_3)\bar{\mathbb{S}}_{s_4}(x_4)\rangle\nonumber\\
	&=(N^2-1)\frac{1}{(4\pi^2)^{4}} 2^{\sum_{l=1}^{4} s_l}i^{\sum_{l=1}^{4} s_l}\nonumber\\
	&\quad\sum_{k_1=0}^{s_1-2}\sum_{k_2=0}^{s_2-2}\sum_{k_3=0}^{s_3-2} \sum_{k_{4} = 0}^{s_{4}-2}{s_1\choose k_1}{s_1\choose k_1+2}   {s_2\choose k_2}{s_2\choose k_2+2}  {s_3\choose k_3}{s_3\choose k_3+2}   {s_{4}\choose k_{4}}{s_{4}\choose k_{4}+2}   \nonumber\\
	&\quad\sum_{\sigma\in P_{2}}\Bigg[(s_{3}-k_{3}+k_{\sigma(1)})!(s_{\sigma(1)}-k_{\sigma(1)}+k_{4})!(s_{4}-k_{4}+k_{\sigma(2)})!(s_{\sigma(2)}-k_{\sigma(2)}+k_{3})!\nonumber\\
	&\quad\frac{(x_{3}-x_{\sigma(1)})_+^{s_{3}-k_{3}+k_{\sigma(1)}}} {\left(\rvert x_{3}-x_{\sigma(1)}\rvert^2\right)^{s_{3}-k_{3}+k_{\sigma(1)}+1}}\frac{(x_{\sigma(1)}-x_{4})_+^{s_{\sigma(1)}-k_{\sigma(1)}+k_{4}}} {\left(\rvert x_{\sigma(1)}-x_{4}\rvert^2\right)^{s_{\sigma(1)}-k_{\sigma(1)}+k_{4}+1}}\nonumber\\
	&\quad\frac{(x_{4}-x_{\sigma(2)})_+^{s_{4}-k_{4}+k_{\sigma(2)}}} {\left(\rvert x_{4}-x_{\sigma(2)}\rvert^2\right)^{s_{4}-k_{4}+k_{\sigma(2)}+1}}\frac{(x_{\sigma(2)}-x_{3})_+^{s_{\sigma(2)}-k_{\sigma(2)}+k_{3}}}{\left(\rvert x_{4}-x_{1}\rvert^2\right)^{s_{4}-k_{4}+k_{1}+1}}\nonumber\\
	&\quad+2(s_{3}-k_{3}+k_{\sigma(1)})!(s_{\sigma(1)}-k_{\sigma(1)}+k_{\sigma(2)})!(s_{\sigma(2)}-k_{\sigma(2)}+k_{4})!(s_{4}-k_{4}+k_{3})!\nonumber\\
	&\quad\frac{(x_{3}-x_{\sigma(1)})_+^{s_{3}-k_{3}+k_{\sigma(1)}}} {\left(\rvert x_{3}-x_{\sigma(1)}\rvert^2\right)^{s_{3}-k_{3}+k_{\sigma(1)}+1}}\frac{(x_{\sigma(1)}-x_{\sigma(2)})_+^{s_{\sigma(1)}-k_{\sigma(1)}+k_{\sigma(2)}}} {\left(\rvert x_{\sigma(1)}-x_{\sigma(2)}\rvert^2\right)^{s_{\sigma(1)}-k_{\sigma(1)}+k_{\sigma(2)}+1}}\nonumber\\
	&\quad\frac{(x_{\sigma(2)}-x_{4})_+^{s_{\sigma(2)}-k_{\sigma(2)}+k_{4}}} {\left(\rvert x_{\sigma(2)}-x_{4}\rvert^2\right)^{s_{\sigma(2)}-k_{\sigma(2)}+k_{4}+1}}\frac{(x_{4}-x_{3})_+^{s_{4}-k_{4}+k_{3}}}{\left(\rvert x_{4}-x_{3}\rvert^2\right)^{s_{4}-k_{4}+k_{3}+1}}\Bigg]
\end{align}
that in the momentum representation reads:
\begin{align}
	&2^4\frac{\Gamma(5)\Gamma(s_1+1)}{\Gamma(3)\Gamma(s_1+3)}
	\frac{\Gamma(5)\Gamma(s_2+1)}{\Gamma(3)\Gamma(s_2+3)}\frac{\Gamma(5)\Gamma(s_3+1)}{\Gamma(3)\Gamma(s_3+3)}\frac{\Gamma(5)\Gamma(s_4+1)}{\Gamma(3)\Gamma(s_4+3)}\nonumber\\
	&\langle\mathbb{O}_{s_1}(p_1)\mathbb{O}_{s_2}(p_2)\mathbb{S}_{s_3}(p_3)\bar{\mathbb{S}}_{s_4}(p_4)\rangle\nonumber\\
	&=(N^2-1)(2\pi)^4\, i^4\,\delta^{(4)}(p_1+p_2+p_3+p_4)\nonumber\\
	&\quad\sum_{k_1=0}^{s_1-2}\sum_{k_2=0}^{s_2-2}\sum_{k_3=0}^{s_3-2} \sum_{k_{4} = 0}^{s_{4}-2}{s_1\choose k_1}{s_1\choose k_1+2}   {s_2\choose k_2}{s_2\choose k_2+2}  {s_3\choose k_3}{s_3\choose k_3+2}   {s_{4}\choose k_{4}}{s_{4}\choose k_{4}+2}   \nonumber\\
	&\quad\sum_{\sigma\in P_{2}}\int \frac{d^4 q}{(2\pi)^4}\Bigg[\frac{(p_{3}+q)_{+}^{s_{3}-k_{3}+k_{\sigma(1)}}}{\rvert p_{3}+q\rvert^2+i\epsilon}\frac{(p_{3}+p_{\sigma(1)}+q)_{+}^{s_{\sigma(1)}-k_{\sigma(1)}+k_{4}}}{\rvert p_{3}+p_{\sigma(1)}+q\rvert^2+i\epsilon}\nonumber\\
	 &\quad\frac{(p_3+p_{\sigma(1)}+p_{4}+q)_{+}^{s_{\sigma(4)}-k_{\sigma(4)}+k_{\sigma(2)}}}{\rvert p_3+p_{\sigma(1)}+p_{4}+q\rvert^2}\frac{(q)_{+}^{s_{\sigma(2)}-k_{\sigma(2)}+k_{3}}}{\rvert q\rvert^2+i\epsilon}\nonumber\\
	 &\quad+2\frac{(p_{3}+q)_{+}^{s_{3}-k_{3}+k_{\sigma(1)}}}{\rvert p_{3}+q\rvert^2+i\epsilon}\frac{(p_{3}+p_{\sigma(1)}+q)_{+}^{s_{\sigma(1)}-k_{\sigma(1)}+k_{\sigma(2)}}}{\rvert p_{3}+p_{\sigma(1)}+q\rvert^2+i\epsilon}\nonumber\\
	 &\quad\frac{(p_3+p_{\sigma(1)}+p_{\sigma(2)}+q)_{+}^{s_{\sigma(2)}-k_{\sigma(2)}+k_{4}}}{\rvert p_3+p_{\sigma(1)}+p_{\sigma(2)}+q\rvert^2+i\epsilon}\frac{(q)_{+}^{s_{4}-k_{4}+k_{3}}}{\rvert q\rvert^2+i\epsilon}\Bigg]
\end{align}
Similarly, we obtain: 
\begin{align}
	&\langle\tilde{\mathbb{O}}_{s_1}(x_1)\tilde{\mathbb{O}}_{s_2}(x_2)\mathbb{S}_{s_3}(x_3)\bar{\mathbb{S}}_{s_4}(x_4)\rangle\nonumber\\
	&= \sum_{k_1=0}^{s_1-2}\sum_{k_2=0}^{s_2-2}\sum_{k_3=0}^{s_3-2}\sum_{k_4=0}^{s_4-2}\frac{\delta}{\delta j_{\tilde{\mathbb{O}}_{s_1k_1}}(x_1)}\frac{\delta}{\delta j_{\tilde{\mathbb{O}}_{s_2k_2}}(x_2)}\frac{\delta}{\delta j_{\mathbb{S}_{s_3k_3}}(x_3)}\frac{\delta}{\delta j_{\bar{\mathbb{S}}_{s_4k_4}}(x_4)} \Gamma_{conf}[j_{\mathbb{O}},j_{\tilde{\mathbb{O}}},j_{\mathbb{S}},j_{\bar{\mathbb{S}}}]\nonumber\\
	&=(N^2-1)\sum_{k_1=0}^{s_1-2}\sum_{k_2=0}^{s_2-2}\sum_{k_3=0}^{s_3-2}\sum_{k_4=0}^{s_4-2}\sum_{\sigma\in P_2}\nonumber\\
	&\quad\Bigg(-\mathcal{D}^{-1}_{s_3k_3,s_{\sigma(1)}k_{\sigma(1)}}(x_3-x_{\sigma(1)}) \mathcal{D}^{-1}_{s_{\sigma(1)}k_{\sigma(1)},s_4k_4}(x_{\sigma(1)}-x_4)\nonumber\\
	&\quad\quad\mathcal{D}^{-1}_{s_4k_4,s_{\sigma(2)}k_{\sigma(2)}}(x_4-x_{\sigma(2)}) \mathcal{D}^{-1}_{s_{\sigma(2)}k_{\sigma(2)},s_3k_3}(x_{\sigma(2)}-x_3) \nonumber\\
	&\quad+2\mathcal{D}^{-1}_{s_3k_3,s_{\sigma(1)}k_{\sigma(1)}}(x_3-x_{\sigma(1)}) \mathcal{D}^{-1}_{s_{\sigma(1)}k_{\sigma(1)},s_{\sigma(2)}k_{\sigma(2)}}(x_{\sigma(1)}-x_{\sigma(2)}) \nonumber\\
	&\quad\quad\mathcal{D}^{-1}_{s_{\sigma(2)}k_{\sigma(2)},s_4k_4}(x_{\sigma(2)}-x_4) \mathcal{D}^{-1}_{s_4k_4,s_3k_3}(x_4-x_3)
	\Bigg)
\end{align}
Explicitly, in the coordinate and momentum representation respectively:
\begin{align}
&2^4\frac{\Gamma(5)\Gamma(s_1+1)}{\Gamma(3)\Gamma(s_1+3)}
\frac{\Gamma(5)\Gamma(s_2+1)}{\Gamma(3)\Gamma(s_2+3)}\frac{\Gamma(5)\Gamma(s_3+1)}{\Gamma(3)\Gamma(s_3+3)}\frac{\Gamma(5)\Gamma(s_4+1)}{\Gamma(3)\Gamma(s_4+3)}\nonumber\\
	&\langle\tilde{\mathbb{O}}_{s_1}(x_1)\tilde{\mathbb{O}}_{s_2}(x_2)\mathbb{S}_{s_3}(x_3)\bar{\mathbb{S}}_{s_4}(x_4)\rangle\nonumber\\
	&=(N^2-1)\frac{1}{(4\pi^2)^{4}} 2^{\sum_{l=1}^{4} s_l}i^{\sum_{l=1}^{4} s_l}\nonumber\\
	&\quad\sum_{k_1=0}^{s_1-2}\sum_{k_2=0}^{s_2-2}\sum_{k_3=0}^{s_3-2} \sum_{k_{4} = 0}^{s_{4}-2}{s_1\choose k_1}{s_1\choose k_1+2}   {s_2\choose k_2}{s_2\choose k_2+2}  {s_3\choose k_3}{s_3\choose k_3+2}   {s_{4}\choose k_{4}}{s_{4}\choose k_{4}+2}   \nonumber\\
	&\quad\sum_{\sigma\in P_{2}}\Bigg[- (s_{3}-k_{3}+k_{\sigma(1)})!(s_{\sigma(1)}-k_{\sigma(1)}+k_{4})!(s_{4}-k_{4}+k_{\sigma(2)})!(s_{\sigma(2)}-k_{\sigma(2)}+k_{3})!\nonumber\\
	&\quad \frac{(x_{3}-x_{\sigma(1)})_+^{s_{3}-k_{3}+k_{\sigma(1)}}} {\left(\rvert x_{3}-x_{\sigma(1)}\rvert^2\right)^{s_{3}-k_{3}+k_{\sigma(1)}+1}}\frac{(x_{\sigma(1)}-x_{4})_+^{s_{\sigma(1)}-k_{\sigma(1)}+k_{4}}} {\left(\rvert x_{\sigma(1)}-x_{4}\rvert^2\right)^{s_{\sigma(1)}-k_{\sigma(1)}+k_{4}+1}}\nonumber\\
	&\quad\frac{(x_{4}-x_{\sigma(2)})_+^{s_{4}-k_{4}+k_{\sigma(2)}}} {\left(\rvert x_{4}-x_{\sigma(2)}\rvert^2\right)^{s_{4}-k_{4}+k_{\sigma(2)}+1}}\frac{(x_{\sigma(2)}-x_{3})_+^{s_{\sigma(2)}-k_{\sigma(2)}+k_{3}}}{\left(\rvert x_{4}-x_{1}\rvert^2\right)^{s_{4}-k_{4}+k_{1}+1}}\nonumber\\
	&\quad+2(s_{3}-k_{3}+k_{\sigma(1)})!(s_{\sigma(1)}-k_{\sigma(1)}+k_{\sigma(2)})!(s_{\sigma(2)}-k_{\sigma(2)}+k_{4})!(s_{4}-k_{4}+k_{3})!\nonumber\\
	&\quad\frac{(x_{3}-x_{\sigma(1)})_+^{s_{3}-k_{3}+k_{\sigma(1)}}} {\left(\rvert x_{3}-x_{\sigma(1)}\rvert^2\right)^{s_{3}-k_{3}+k_{\sigma(1)}+1}}\frac{(x_{\sigma(1)}-x_{\sigma(2)})_+^{s_{\sigma(1)}-k_{\sigma(1)}+k_{\sigma(2)}}} {\left(\rvert x_{\sigma(1)}-x_{\sigma(2)}\rvert^2\right)^{s_{\sigma(1)}-k_{\sigma(1)}+k_{\sigma(2)}+1}}\nonumber\\
	&\quad\frac{(x_{\sigma(2)}-x_{4})_+^{s_{\sigma(2)}-k_{\sigma(2)}+k_{4}}} {\left(\rvert x_{\sigma(2)}-x_{4}\rvert^2\right)^{s_{\sigma(2)}-k_{\sigma(2)}+k_{4}+1}}\frac{(x_{4}-x_{3})_+^{s_{4}-k_{4}+k_{3}}}{\left(\rvert x_{4}-x_{3}\rvert^2\right)^{s_{4}-k_{4}+k_{3}+1}}\Bigg]
\end{align}
\begin{align}
	&2^4\frac{\Gamma(5)\Gamma(s_1+1)}{\Gamma(3)\Gamma(s_1+3)}
	\frac{\Gamma(5)\Gamma(s_2+1)}{\Gamma(3)\Gamma(s_2+3)}\frac{\Gamma(5)\Gamma(s_3+1)}{\Gamma(3)\Gamma(s_3+3)}\frac{\Gamma(5)\Gamma(s_4+1)}{\Gamma(3)\Gamma(s_4+3)}\nonumber\\
	&\langle\tilde{\mathbb{O}}_{s_1}(p_1)\tilde{\mathbb{O}}_{s_2}(p_2)\mathbb{S}_{s_3}(p_3)\bar{\mathbb{S}}_{s_4}(p_4)\rangle\rangle\nonumber\\
	&=(N^2-1)(2\pi)^4\, i^4\,\delta^{(4)}(p_1+p_2+p_3+p_4)\nonumber\\
	&\quad\sum_{k_1=0}^{s_1-2}\sum_{k_2=0}^{s_2-2}\sum_{k_3=0}^{s_3-2} \sum_{k_{4} = 0}^{s_{4}-2}{s_1\choose k_1}{s_1\choose k_1+2}   {s_2\choose k_2}{s_2\choose k_2+2}  {s_3\choose k_3}{s_3\choose k_3+2}   {s_{4}\choose k_{4}}{s_{4}\choose k_{4}+2}   \nonumber\\
	&\quad\sum_{\sigma\in P_{2}}\int \frac{d^4 q}{(2\pi)^4}\Bigg[-\frac{(p_{3}+q)_{+}^{s_{3}-k_{3}+k_{\sigma(1)}}}{\rvert p_{3}+q\rvert^2+i\epsilon}\frac{(p_{3}+p_{\sigma(1)}+q)_{+}^{s_{\sigma(1)}-k_{\sigma(1)}+k_{4}}}{\rvert p_{3}+p_{\sigma(1)}+q\rvert^2+i\epsilon}\nonumber\\
	&\quad\frac{(p_3+p_{\sigma(1)}+p_{4}+q)_{+}^{s_{\sigma(4)}-k_{\sigma(4)}+k_{\sigma(2)}}}{\rvert p_3+p_{\sigma(1)}+p_{4}+q\rvert^2}\frac{(q)_{+}^{s_{\sigma(2)}-k_{\sigma(2)}+k_{3}}}{\rvert q\rvert^2+i\epsilon}\nonumber\\
	&\quad+2\frac{(p_{3}+q)_{+}^{s_{3}-k_{3}+k_{\sigma(1)}}}{\rvert p_{3}+q\rvert^2+i\epsilon}\frac{(p_{3}+p_{\sigma(1)}+q)_{+}^{s_{\sigma(1)}-k_{\sigma(1)}+k_{\sigma(2)}}}{\rvert p_{3}+p_{\sigma(1)}+q\rvert^2+i\epsilon}\nonumber\\
	&\quad\frac{(p_3+p_{\sigma(1)}+p_{\sigma(2)}+q)_{+}^{s_{\sigma(2)}-k_{\sigma(2)}+k_{4}}}{\rvert p_3+p_{\sigma(1)}+p_{\sigma(2)}+q\rvert^2+i\epsilon}\frac{(q)_{+}^{s_{4}-k_{4}+k_{3}}}{\rvert q\rvert^2+i\epsilon}\Bigg]
\end{align}

\subsection{$4$-point correlators in Euclidean space-time}

In Euclidean space-time we obtain as well:
\begin{align}
	&2^4\frac{\Gamma(5)\Gamma(s_1+1)}{\Gamma(3)\Gamma(s_1+3)}
	\frac{\Gamma(5)\Gamma(s_2+1)}{\Gamma(3)\Gamma(s_2+3)}\frac{\Gamma(5)\Gamma(s_3+1)}{\Gamma(3)\Gamma(s_3+3)}\frac{\Gamma(5)\Gamma(s_4+1)}{\Gamma(3)\Gamma(s_4+3)}\nonumber\\
	&\langle\mathbb{O}^E_{s_1}(x_1)\mathbb{O}^E_{s_2}(x_2)\mathbb{S}^E_{s_3}(x_3)\bar{\mathbb{S}}^E_{s_4}(x_4)\rangle\nonumber\\
	&=(N^2-1)\frac{1}{(4\pi^2)^{4}}2^{\sum_{l=1}^{4} s_l}(-1)^{\sum_{l=1}^{4} s_l}\nonumber\\
	&\quad\sum_{k_1=0}^{s_1-2}\sum_{k_2=0}^{s_2-2}\sum_{k_3=0}^{s_3-2} \sum_{k_{4} = 0}^{s_{4}-2}{s_1\choose k_1}{s_1\choose k_1+2}   {s_2\choose k_2}{s_2\choose k_2+2}  {s_3\choose k_3}{s_3\choose k_3+2}   {s_{4}\choose k_{4}}{s_{4}\choose k_{4}+2}   \nonumber\\
	&\quad\sum_{\sigma\in P_{2}}\Bigg[(s_{3}-k_{3}+k_{\sigma(1)})!(s_{\sigma(1)}-k_{\sigma(1)}+k_{4})!(s_{4}-k_{4}+k_{\sigma(2)})!(s_{\sigma(2)}-k_{\sigma(2)}+k_{3})!\nonumber\\
	&\quad\frac{(x_{3}-x_{\sigma(1)})_z^{s_{3}-k_{3}+k_{\sigma(1)}}} {\left(( x_{3}-x_{\sigma(1)})^2\right)^{s_{3}-k_{3}+k_{\sigma(1)}+1}}\frac{(x_{\sigma(1)}-x_{4})_z^{s_{\sigma(1)}-k_{\sigma(1)}+k_{4}}} {\left(( x_{\sigma(1)}-x_{4})^2\right)^{s_{\sigma(1)}-k_{\sigma(1)}+k_{4}+1}}\nonumber\\
	&\quad\frac{(x_{4}-x_{\sigma(2)})_z^{s_{4}-k_{4}+k_{\sigma(2)}}} {\left(( x_{4}-x_{\sigma(2)})^2\right)^{s_{4}-k_{4}+k_{\sigma(2)}+1}}\frac{(x_{\sigma(2)}-x_{3})_z^{s_{\sigma(2)}-k_{\sigma(2)}+k_{3}}}{\left(( x_{4}-x_{1})^2\right)^{s_{4}-k_{4}+k_{1}+1}}\nonumber\\
	&\quad+2(s_{3}-k_{3}+k_{\sigma(1)})!(s_{\sigma(1)}-k_{\sigma(1)}+k_{\sigma(2)})!(s_{\sigma(2)}-k_{\sigma(2)}+k_{4})!(s_{4}-k_{4}+k_{3})!\nonumber\\
	&\quad\frac{(x_{3}-x_{\sigma(1)})_z^{s_{3}-k_{3}+k_{\sigma(1)}}} {\left(( x_{3}-x_{\sigma(1)})^2\right)^{s_{3}-k_{3}+k_{\sigma(1)}+1}}\frac{(x_{\sigma(1)}-x_{\sigma(2)})_z^{s_{\sigma(1)}-k_{\sigma(1)}+k_{\sigma(2)}}} {\left(( x_{\sigma(1)}-x_{\sigma(2)})^2\right)^{s_{\sigma(1)}-k_{\sigma(1)}+k_{\sigma(2)}+1}}\nonumber\\
	&\quad\frac{(x_{\sigma(2)}-x_{4})_z^{s_{\sigma(2)}-k_{\sigma(2)}+k_{4}}} {\left(( x_{\sigma(2)}-x_{4})^2\right)^{s_{\sigma(2)}-k_{\sigma(2)}+k_{4}+1}}\frac{(x_{4}-x_{3})_z^{s_{4}-k_{4}+k_{3}}}{\left(( x_{4}-x_{3})^2\right)^{s_{4}-k_{4}+k_{3}+1}}\Bigg]
\end{align}
and:
\begin{align}
	&2^4\frac{\Gamma(5)\Gamma(s_1+1)}{\Gamma(3)\Gamma(s_1+3)}
	\frac{\Gamma(5)\Gamma(s_2+1)}{\Gamma(3)\Gamma(s_2+3)}\frac{\Gamma(5)\Gamma(s_3+1)}{\Gamma(3)\Gamma(s_3+3)}\frac{\Gamma(5)\Gamma(s_4+1)}{\Gamma(3)\Gamma(s_4+3)}\nonumber\\
	&\langle\tilde{\mathbb{O}}^E_{s_1}(x_1)\tilde{\mathbb{O}}^E_{s_2}(x_2)\mathbb{S}^E_{s_3}(x_3)\bar{\mathbb{S}}^E_{s_4}(x_4)\rangle\nonumber\\
	&=(N^2-1)\frac{1}{(4\pi^2)^{4}}2^{\sum_{l=1}^{4} s_l}(-1)^{\sum_{l=1}^{4} s_l}\nonumber\\
	&\quad\sum_{k_1=0}^{s_1-2}\sum_{k_2=0}^{s_2-2}\sum_{k_3=0}^{s_3-2} \sum_{k_{4} = 0}^{s_{4}-2}{s_1\choose k_1}{s_1\choose k_1+2}   {s_2\choose k_2}{s_2\choose k_2+2}  {s_3\choose k_3}{s_3\choose k_3+2}   {s_{4}\choose k_{4}}{s_{4}\choose k_{4}+2}   \nonumber\\
	&\quad\sum_{\sigma\in P_{2}}\Bigg[- (s_{3}-k_{3}+k_{\sigma(1)})!(s_{\sigma(1)}-k_{\sigma(1)}+k_{4})!(s_{4}-k_{4}+k_{\sigma(2)})!(s_{\sigma(2)}-k_{\sigma(2)}+k_{3})!\nonumber\\
	&\quad \frac{(x_{3}-x_{\sigma(1)})_z^{s_{3}-k_{3}+k_{\sigma(1)}}} {\left(( x_{3}-x_{\sigma(1)})^2\right)^{s_{3}-k_{3}+k_{\sigma(1)}+1}}\frac{(x_{\sigma(1)}-x_{4})_z^{s_{\sigma(1)}-k_{\sigma(1)}+k_{4}}} {\left(( x_{\sigma(1)}-x_{4})^2\right)^{s_{\sigma(1)}-k_{\sigma(1)}+k_{4}+1}}\nonumber\\
	&\quad\frac{(x_{4}-x_{\sigma(2)})_z^{s_{4}-k_{4}+k_{\sigma(2)}}} {\left(( x_{4}-x_{\sigma(2)})^2\right)^{s_{4}-k_{4}+k_{\sigma(2)}+1}}\frac{(x_{\sigma(2)}-x_{3})_z^{s_{\sigma(2)}-k_{\sigma(2)}+k_{3}}}{\left(( x_{4}-x_{1})^2\right)^{s_{4}-k_{4}+k_{1}+1}}\nonumber\\
	&\quad+2(s_{3}-k_{3}+k_{\sigma(1)})!(s_{\sigma(1)}-k_{\sigma(1)}+k_{\sigma(2)})!(s_{\sigma(2)}-k_{\sigma(2)}+k_{4})!(s_{4}-k_{4}+k_{3})!\nonumber\\
	&\quad\frac{(x_{3}-x_{\sigma(1)})_z^{s_{3}-k_{3}+k_{\sigma(1)}}} {\left(( x_{3}-x_{\sigma(1)})^2\right)^{s_{3}-k_{3}+k_{\sigma(1)}+1}}\frac{(x_{\sigma(1)}-x_{\sigma(2)})_z^{s_{\sigma(1)}-k_{\sigma(1)}+k_{\sigma(2)}}} {\left(( x_{\sigma(1)}-x_{\sigma(2)})^2\right)^{s_{\sigma(1)}-k_{\sigma(1)}+k_{\sigma(2)}+1}}\nonumber\\
	&\quad\frac{(x_{\sigma(2)}-x_{4})_z^{s_{\sigma(2)}-k_{\sigma(2)}+k_{4}}} {\left(( x_{\sigma(2)}-x_{4})^2\right)^{s_{\sigma(2)}-k_{\sigma(2)}+k_{4}+1}}\frac{(x_{4}-x_{3})_z^{s_{4}-k_{4}+k_{3}}}{\left(( x_{4}-x_{3})^2\right)^{s_{4}-k_{4}+k_{3}+1}}\Bigg]
\end{align}

\section{UV asymptotics of Euclidean $n$-point correlators}  \label{s0}

\subsection{Operator mixing}

The Euclidean $n$-point correlator of local renormalized operators $\mathcal{O}_i(x)$ that mix under renormalization:
\begin{equation}\label{key}
	\langle \mathcal{O}_{k_1}(x_1)\ldots \mathcal{O}_{k_n}(x_n) \rangle = G^{(n)}_{k_1 \ldots k_n}( x_1,\ldots,  x_n; \mu, g(\mu))
\end{equation}
satisfies the Callan-Symanzik equation:
\begin{equation}\label{CallanSymanzik}
	\begin{split}
		& \Big(\sum_{\alpha = 1}^n x_\alpha \cdot \frac{\partial}{\partial x_\alpha} + \beta(g)\frac{\partial}{\partial g} + \sum_{\alpha = 1}^n D_{\mathcal{O}_\alpha}\Big)G^{(n)}_{k_1 \ldots k_n} + \\
		& \sum_a \Big(\gamma_{k_1a}(g) G^{(n)}_{ak_2 \ldots k_n} + \gamma_{k_2a}(g) G^{(n)}_{k_1 a k_3 \ldots k_n} \cdots +\gamma_{k_n a}(g) G^{(n)}_{k_1 \ldots a}\Big) = 0
	\end{split}
\end{equation}
where $D_{\mathcal{O}_i}$ is the canonical dimension of $\mathcal{O}_i(x)$ and $\gamma_{ij}(g)$ the matrix of the anomalous dimensions.	
Its solution reads:
\begin{equation}\label{csformula}
	\begin{split}
		&G^{(n)}_{k_1 \ldots k_n}(\lambda x_1,\ldots, \lambda x_n; \mu, g(\mu))  \\
		&= \sum_{j_1 \ldots j_n} Z_{k_1 j_1} (\lambda)\ldots Z_{k_n j_n}(\lambda)\hspace{0.1cm} \lambda^{-\sum_{i=1}^nD_{\mathcal{O}_i}}G^{(n)}_{j_1 \ldots j_n }( x_1, \ldots, x_n; \mu, g\Big(\frac{\mu}{\lambda}\Big))
	\end{split}
\end{equation}
where in matrix notation:
\begin{equation}\label{eqZ}
	\Bigg(\frac{\partial}{\partial g} + \frac{\gamma(g)}{\beta(g)}\Bigg)Z(\lambda) = 0
\end{equation}
with $g\equiv g(\mu)$ and:
\begin{equation} \label{ZZ}
	Z(\lambda) = P\exp \Big(\int_{g(\mu)}^{g(\frac{\mu}{\lambda})}\frac{\gamma(g')}{\beta(g')} dg'\Big)
\end{equation}
Hence, the corresponding UV asymptotics as $\lambda \rightarrow 0$ reads:
\begin{equation}\label{csformula0}
	\begin{split}
		&G^{(n)}_{k_1 \ldots k_n}(\lambda x_1,\ldots, \lambda x_n; \mu, g(\mu))  \\
		&\sim \sum_{j_1 \ldots j_n} Z_{k_1 j_1} (\lambda)\ldots Z_{k_n j_n}(\lambda)\hspace{0.1cm}  \lambda^{-\sum_{i=1}^nD_{\mathcal{O}_i}}G^{(n)}_{conf\, j_1 \ldots j_n }( x_1, \ldots, x_n)
	\end{split}
\end{equation}
provided that $G^{(n)}_{conf\, j_1 \ldots j_n }( x_1, \ldots, x_n)$ -- the conformal correlators to the lowest order of perturbation theory -- do not vanish. \par
The evaluation of the asymptotics above involves the estimate of each term in eq. \eqref{csformula0} that in turn involves the computation of the path-ordered exponential defining $Z_{ij}(\lambda)$ in eq. \eqref{ZZ}. Both the computations are technically challenging, even in the special case where $\gamma(g)$ is triangular so that the expansion of the path-ordered exponential in eq. \eqref{ZZ} terminates to a finite order \cite{BB}.\par
Therefore, it is of the outmost importance to establish whether a renormalization scheme 
exists where $Z(\lambda)$ is diagonalizable to all perturbative orders. Indeed, in such a scheme the sum in eq. \eqref{csformula0} would reduce to just one term and the path-ordered exponential to just the ordinary exponential, as in the multiplicatively renormalizable case.

\subsection{Nonresonant diagonal renormalization scheme}

The question above has been addressed on the basis of the differential-geometric interpretation \cite{MB1} of the operator mixing in massless QCD-like theories to all orders of perturbation theory that we summarize in the following. We interpret a finite change of the basis of renormalized operators in matrix notation:
\begin{equation}\label{linearcomb}
	\mathcal{O}'(x) = S(g) \mathcal{O}(x)
\end{equation}
as a formal real-analytic invertible gauge transformation $S(g)$ that depends on $g$ \cite{MB1}. Under the action of the aforementioned gauge transformation, the matrix:
\begin{equation}
	A(g) = -\frac{\gamma(g)}{\beta(g)} = \frac{1}{g} \Big(\frac{\gamma_0}{\beta_0} + \cdots\Big)
\end{equation}
associated to the differential equation for $Z(\lambda)$:
\begin{equation}
	\Big(\frac{\partial}{\partial g} - A(g)\Big) Z(\lambda) = 0
\end{equation}
defines a connection $A(g)$:
\begin{eqnarray} \label{sys2}
A(g)= \frac{1}{g} \left(A_0 + \sum^{\infty}_ {n=1} A_{2n} g^{2n} \right)
\end{eqnarray}
with a regular singularity at $g = 0$ that transforms as:
\begin{equation}
	A'(g) = S(g)A(g)S^{-1}(g) + \frac{\partial S(g)}{\partial g} S^{-1}(g)
\end{equation}
with:
\begin{equation}
	\mathcal{D} = \frac{\partial }{\partial g} - A(g)
\end{equation}
the corresponding covariant derivative.
Consequently, $Z(\lambda)$ is interpreted as a Wilson line that transforms as:
\begin{equation}
	Z'(\lambda) = S(g(\mu))Z(\lambda)S^{-1}(g(\frac{\mu}{\lambda}))
\end{equation}
for the gauge transformation $S(g)$.	Moreover,	everything that we have mentioned also applies by allowing the coupling $g$ to be complex valued. It follows from the Poincar\'e-Dulac theorem that \cite{MB1}, if any two eigenvalues $\lambda_1, \lambda_2, \ldots$ of the matrix $\frac{\gamma_0}{\beta_0}$, in nonincreasing order $\lambda_1 \geq \lambda_2 \geq \ldots$, do not differ by a positive even integer:
\begin{equation} \label{nr}
	\lambda_i - \lambda_j - 2k \neq 0
\end{equation}
for $i \leq j$ and $k$ a positive integer, then a formal holomorphic gauge transformation exists that sets $A(g)$ in the canonical nonresonant form \cite{MB1}:
\begin{equation} \label{1loop}
	A'(g) = \frac{\gamma_0}{\beta_0}\frac{1}{g}
\end{equation}
that is one-loop exact to all orders of perturbation theory. 
As a consequence, if in addition $\frac{\gamma_0}{\beta_0}$ is diagonalizable, $Z(\lambda)$ is diagonalizable as well and its eigenvalues $Z_{\mathcal{O}_i}(\lambda)$ are computed by \cite{MB1}:
\begin{equation}
	Z_{\mathcal{O}_i}(\lambda) = \Bigg(\frac{g(\mu)}{g(\frac{\mu}{\lambda})}\Bigg)^{\frac{\gamma_{0\mathcal{O}_i}}{\beta_0}} 
\end{equation}
with $\gamma_{0\mathcal{O}_i}$ the eigenvalues of $\gamma_0$.
Correspondingly, in the nonresonant diagonal scheme:
\begin{equation} \label{eqrg}
	\langle \mathcal{O}_{j_1}(\lambda x_1)\ldots\mathcal{O}_{j_n}(\lambda x_n)\rangle_{conn} \sim \frac{Z_{\mathcal{O}_{j_1}}(\lambda)\ldots Z_{\mathcal{O}_{j_n}}(\lambda)}{\lambda^{D_{\mathcal{O}_1}+\cdots+D_{\mathcal{O}_n}}} G^{(n)}_{conf\,j_1 \ldots j_n }( x_1, \ldots, x_n)
\end{equation}
provided that  $G^{(n)}_{conf\,j_1 \ldots j_n }( x_1, \ldots, x_n)$ does not vanish, with:
\begin{eqnarray} 
	g^2(\frac{\mu}{\lambda})
	\sim \dfrac{1}{\beta_0 \log(\frac{1}{\lambda^2})}\left(1-\frac{\beta_1}{\beta_0^2}\frac{\log\log(\frac{1}{\lambda^2})}{\log(\frac{1}{\lambda^2})}\right) \nonumber \\
\end{eqnarray}
as $\lambda \rightarrow 0$.\par
To make the present paper self-contained we provide the construction \cite{MB1} of the nonresonant diagonal scheme to all orders of perturbation theory under the above assumptions.\par
The construction proceeds by induction on $k=1,2, \cdots$ by demonstrating that, once $A_0$ and the first $k-1$ matrix coefficients $A_2,\cdots,A_{2(k-1)}$ in eq. \eqref{sys2} have been set in the canonical nonresonant form in eq. \eqref{1loop} -- i.e., $A_0$ diagonal and $ A_2,\cdots,A_{2(k-1)}=0$ --, a formal holomorphic gauge transformation exists that leaves them invariant and also sets the $k$-th coefficient $A_{2k}$ to $0$. \par
The $0$ step of the induction consists just in setting $A_0$ in diagonal form -- with the eigenvalues in nonincreasing order -- by a global (i.e., constant) gauge transformation. \par
At the $k$-th step we choose the holomorphic gauge transformation in the form: 
\begin{eqnarray}
S_k(g)=1+ g^{2k} H_{2k}
\end{eqnarray}
with $H_{2k}$ a matrix to be found momentarily. Its inverse is:
\begin{eqnarray}
S^{-1}_k(g)= (1+ g^{2k} H_{2k})^{-1} = 1- g^{2k} H_{2k} + \cdots
\end{eqnarray}
where the dots represent terms of order higher than $g^{2k}$.
The gauge action of $S_k(g)$ on the connection $A(g)$ furnishes:
\begin{eqnarray} \label{ind}
A'(g) &= & 2k g^{2k-1} H_{2k} ( 1- g^{2k} H_{2k})^{-1}+  (1+ g^{2k} H_{2k}) A(g)( 1- g^{2k} H_{2k})^{-1} \nonumber \\
&= & 2k g^{2k-1} H_{2k} ( 1- g^{2k} H_{2k})^{-1} +  (1+ g^{2k} H_{2k})  \frac{1}{g} \left(A_0 + \sum^{\infty}_ {n=1} A_{2n} g^{2n} \right)  ( 1- g^{2k} H_{2k})^{-1} \nonumber \\
&= & 2k g^{2k-1} H_{2k} ( 1- \cdots) +  (1+ g^{2k} H_{2k})  \frac{1}{g} \left(A_0 + \sum^{\infty}_ {n=1} A_{2n} g^{2n} \right)  ( 1- g^{2k} H_{2k}+\cdots) \nonumber \\
&= & 2k g^{2k-1} H_{2k}  +    \frac{1}{g} \left(A_0 + \sum^k_ {n=1} A_{2n} g^{2n} \right) + g^{2k-1} (H_{2k}A_0-A_0H_{2k}) + \cdots\nonumber \\
&= &  g^{2k-1} (2k H_{2k} + H_{2k} A_0 - A_0 H_{2k})  + A_{2(k-1)}(g) + g^{2k-1} A_{2k}+ \cdots\nonumber \\
\end{eqnarray}
where we have skipped in the dots all the terms that contribute to an order higher than $g^{2k-1}$, and we have set:
\begin{eqnarray}
A_{2(k-1)}(g) =  \frac{1}{g} \left(A_0 + \sum^{k-1}_ {n=1} A_{2n} g^{2n} \right)
\end{eqnarray}
that is the part of $A(g)$ that is not affected by the gauge transformation $S_k(g)$, and thus verifies the hypotheses of the induction --  i.e., that $A_2, \cdots, A_{2(k-1)}$ vanish. \par
Therefore, by eq. \eqref{ind} the $k$-th matrix coefficient $A_{2k}$ may be eliminated from the expansion of $A'(g)$ to the order $g^{2k-1}$ provided that an $H_{2k}$ exists such that:
\begin{eqnarray}
A_{2k}+(2k H_{2k} + H_{2k} A_0 - A_0 H_{2k})= A_{2k}+ (2k-ad_{A_0}) H_{2k}=0
\end{eqnarray}
with $ad_{A_0}Y=[A_0,Y]$.\par
If the inverse of $ad_{A_0}-2k$ exists, the unique solution for $H_{2k}$ is:
\begin{eqnarray}
H_{2k}=(ad_{A_0}-2k)^{-1} A_{2k}
\end{eqnarray}
Hence, to complete the induction, we should demonstrate that, under the above assumptions, $ad_{A_0}-2k$ is invertible, i.e., its kernel is trivial. 
\par
Now $ad_{\Lambda}-2k$, as a linear operator that acts on matrices, is diagonal, with eigenvalues $\lambda_{i}-\lambda_{j}-2k$ and the matrices $E_{ij}$, whose only nonvanishing entries are $(E_{ij})_{ij}$, as eigenvectors.\par The eigenvectors $E_{ij}$, normalized in such a way that $(E_{ij})_{ij}=1$, form an orthonormal basis for the matrices. 
Thus, $E_{ij}$ belongs to the kernel of $ad_{\Lambda}-2k$ if and only if $\lambda_{i}-\lambda_{j}-2k=0$. As a consequence, since  $\lambda_{i}-\lambda_{j}-2k \neq 0$ for every $i,j$ by the above assumptions, the kernel of $ad_{\Lambda}-2k$ only contains $0$, and the construction is complete.

  \subsection{Anomalous dimensions of twist-$2$ operators}

We define the bare operators with $s \geq 2$ and $k \geq 0$ \cite{Belitsky:1998gc}:
   \begin{equation}
 \mathcal{O}^{(k)}_{Bs} = (-i\partial_+)^{k}\mathcal{O}_{Bs}
   \end{equation}
 that, to the leading order of perturbation theory, for $k>0$ are conformal descendants \cite{Braun:2003rp} of the corresponding primary conformal operator 
 $\mathcal{O}^{(0)}_{Bs}=\mathcal{O}_{Bs}$ in the standard basis.
 As a consequence of the operator mixing we obtain \cite{Belitsky:1998gc,Braun:2003rp} for the renormalized operators:
  \begin{equation}\label{m}
   \mathcal{O}^{(k)}_{s}= \sum^{s}_{i} Z_{si} \mathcal{O}^{(k+s-i)}_{B i}
  \end{equation}
  with the mixing matrix $Z$ and the matrix of the anomalous dimensions:
  \begin{equation}
   \gamma(g) =- \frac{\partial Z}{\partial \log \mu} Z^{-1}=\sum_{j=0}^{\infty}\gamma_j \, g^{2j+2}
   \end{equation}
  lower triangular and $\gamma_0$ diagonal in the $\overline{MS}$ scheme \cite{Belitsky:1998gc,Braun:2003rp}.
  In our notation the eigenvalues of $\gamma_0$ are given by \cite{Belitsky:1998gc,Belitsky:2004sc}:
   \begin{equation}
 \label{an1}
\gamma_{0s}= \frac{2}{(4 \pi)^2} \left(4 \psi(s+1) - 4 \psi(1) -\frac{11}{3} - 8 \frac{s^2+s+1}{(s-1)s(s+1)(s+2)}\right)
  \end{equation}
  for $\mathbb{O}^{(k)}_{s}$ with even $s\geq 2$, and:
  \begin{equation}
  \label{an2}
\gamma_{0s}= \frac{2}{(4 \pi)^2}  \left(4 \psi(s+1) - 4 \psi(1) -\frac{11}{3} - 8 \frac{s^2+s-2}{(s-1)s(s+1)(s+2)}  \right)
  \end{equation}
  for $\tilde{\mathbb{O}}^{(k)}_{s}$ with odd $s\geq 3$. $\gamma_{02}=0$ consistently with the conservation of the stress-energy tensor.
  The eigenvalues of $\gamma_0$ for $\mathbb{S}^{(k)}_{s}$ and $\bar{\mathbb{S}}^{(k)}_{s}$ are given by \cite{Belitsky:2004sc}:
\begin{equation}
	\gamma_{0s}=  \frac{2}{(4 \pi)^2} \left(4\psi(s+1)-4\psi(1)-\frac{11}{3}\right)
\end{equation}
 with even $s\geq 2$.\par
Interestingly, from the above equations it follows that in $SU(N)$ YM theory $\gamma_0$ is independent of $N$ and thus has no nonplanar contribution \cite{Aglietti:2021bem}. \par 
Moreover, we have verified numerically that the nonresonant condition in eq. \eqref{nr} is satisfied for all the collinear twist-$2$ operators in the standard basis up to $s=10^4$. \par
Hence, the nonresonant diagonal basis exists for the collinear twist-$2$ operators and restricts to the lowest order of perturbation theory to the standard basis because $\gamma_0$ is diagonal in this basis. \par
Therefore, the standard basis may be employed to compute the conformal correlators in eq. \eqref{eqrg}.\par

  \section{Generating functional of Euclidean RG-improved correlators \label{s6}}

From the vantage point of view of the 't Hooft large-$N$ expansion it is convenient to employ the rescaled operators in eqs. \eqref{rescale} in Minkowskian and Euclidean space-time respectively, so that the $2$-point correlators are of order $1$ for large $N$. 
By choosing again the operators $\mathbb{O}'_{s}(x)$ as example, the corresponding conformal correlators acquire the simpler form in Minkowskian:
  \begin{align}
  	\label{standardn}
  	\nonumber
  	&\langle \mathbb{O}'_{s_1}(x_1)\ldots \mathbb{O}'_{s_n}(x_n)\rangle_{conn}\nonumber\\\nonumber &=(N^{2-n}-N^{-n})\frac{1}{(4\pi^2)^n}2^{\sum_{l=1}^ns_l}i^{\sum_{l=1}^ns_l}\sum_{k_1=0}^{s_1-2}\ldots \sum_{k_n = 0}^{s_n-2}{s_1\choose k_1}{s_1\choose k_1+2}\ldots {s_n\choose k_n}{s_n\choose k_n+2}\\\nonumber
  	&\quad\frac{(-1)^n}{n}\sum_{\sigma\in P_n}(s_{\sigma(1)}-k_{\sigma(1)}+k_{\sigma(2)})!\ldots(s_{\sigma(n)}-k_{\sigma(n)}+k_{\sigma(1)})!\\
  	&\quad\frac{(x_{\sigma(1)}-x_{\sigma(2)})_+^{s_{\sigma(1)}-k_{\sigma(1)}+k_{\sigma(2)}}}{\left(\rvert x_{\sigma(1)}-x_{\sigma(2)}\rvert^2\right)^{s_{\sigma(1)}-k_{\sigma(1)}+k_{\sigma(2)}+1}}
  	\cdots\frac{(x_{\sigma(n)}-x_{\sigma(1)})_+^{s_{\sigma(n)}-k_{\sigma(n)}+k_{\sigma(1)}}}{\left(\rvert x_{\sigma(n)}-x_{\sigma(1)}\rvert^2\right)^{s_{\sigma(n)}-k_{\sigma(n)}+k_{\sigma(1)}+1}}
  \end{align}
and Euclidean space-time:
\begin{align} \label{OE}
	&\langle \mathbb{O'}^E_{s_1}(x_1)\ldots \mathbb{O'}^E_{s_n}(x_n)\rangle_{conn}\nonumber\\
	&=(N^{2-n}-N^{-n})\frac{1}{(4\pi^2)^n}2^{\sum_{l=1}^ns_l}(-1)^{\sum_{l=1}^ns_l}\sum_{k_1=0}^{s_1-2}\ldots \sum_{k_n = 0}^{s_n-2}{s_1\choose k_1}{s_1\choose k_1+2}\ldots {s_n\choose k_n}{s_n\choose k_n+2}\nonumber\\
	&\quad\frac{1}{n}\sum_{\sigma\in P_{n}}(s_{\sigma(1)}-k_{\sigma(1)}+k_{\sigma(2)})!\ldots(s_{\sigma(n)}-k_{\sigma(n)}+k_{\sigma(1)})!\nonumber\\
	&\quad\frac{(x_{\sigma(1)}-x_{\sigma(2)})_{z}^{s_{\sigma(1)}-k_{\sigma(1)}+k_{\sigma(2)}}}{\left(( x_{\sigma(1)}-x_{\sigma(2)})^2\right)^{s_{\sigma(1)}-k_{\sigma(1)}+k_{\sigma(2)}+1}}
	\cdots\frac{(x_{\sigma(n)}-x_{\sigma(1)})_{z}^{s_{\sigma(n)}-k_{\sigma(n)}+k_{\sigma(1)}}}{\left(( x_{\sigma(n)}-x_{\sigma(1)})^2\right)^{s_{\sigma(n)}-k_{\sigma(n)}+k_{\sigma(1)}+1}}\,
\end{align}
according to eqs. \eqref{wexplicit} and \eqref{Wick}.
Correspondingly, the generating functional of the Euclidean $n$-point correlators of twist-$2$ operators decomposes into its planar $\mathcal{W}^E_{sphere}$ and LO-nonplanar $\mathcal{W}^E_{torus}$ contributions.\par
Perturbatively, according to general principles of the 't Hooft large-$N$ expansion \cite{tHooft:1973alw}, $\mathcal{W}^E_{torus}$ involves a sum of  Feynman diagrams that have the topology of a punctured torus in the 't Hooft double-line representation.\par
Nonperturbatively, $\mathcal{W}^E_{torus}$ must involve the sum of the glueball one-loop diagrams and it has been predicted in \cite{Bochicchio:2016toi} -- on the basis of the existence of a large-$N$ nonperturbative effective action for the glueballs -- that it should have the structure of the logarithm of a functional determinant. \par
The UV asymptotics as $\lambda \rightarrow 0$ of $\mathcal{W}^E_{sphere}$ and $\mathcal{W}^E_{torus}$ for twist-$2$ operators in the nonresonant diagonal 
scheme follows from the computation of the RG-improved correlators (section \ref{s0}):
\begin{equation}
\mathcal{W}^E_{sphere}[j_{\mathcal{O'}^E},\lambda] \sim \mathcal{W}^E_{asym \, sphere}[j_{\mathcal{O'}^E},\lambda]
\end{equation}
and:
\begin{equation}
\mathcal{W}^E_{torus}[j_{\mathcal{O'}^E},\lambda] \sim \mathcal{W}^E_{asym \, torus }[j_{\mathcal{O'}^E},\lambda]
\end{equation}
with:
\begin{equation}
\mathcal{W}^E_{asym \, sphere}[j_{\mathcal{O'}^E},\lambda]=-N^2 \mathcal{W}^E_{asym \, torus}[j_{\mathcal{O'}^E},\lambda]\\
\end{equation}
according to eq. \eqref{wexplicit}, where:
  \begin{align}
  		\label{fullGenEexpZ}
	&\mathcal{W}^E_{asym \, torus}[J_{\mathbb{O}^{'E}},J_{\tilde{\mathbb{O}}^{'E}},J_{\mathbb{S}^{'E}},J_{\bar{\mathbb{S}}^{'E}},\lambda]=\nonumber\\
	&+\frac{1}{2}\log\Det\left(I+\sum_{k=0}^{s-2}{s\choose k}{s\choose k+2}(-\overrightarrow{\partial}_z)^{s-k-1}\Laplace^{-1}\frac{(Z_{\mathbb{O'}_{s}}(\lambda) J_{\mathbb{O}^{'E}_{s}}+Z_{\tilde{\mathbb{O}'}_{s}}(\lambda)J_{\tilde{\mathbb{O}}^{'E}_{s}})}{N\,\lambda^{2+s}}(-\overrightarrow{\partial}_z)^{k+1} \right)\nonumber\\
	&+\frac{1}{2}\log\Det\left(I+\sum_{k=0}^{s-2}{s\choose k}{s\choose k+2}(-\overrightarrow{\partial}_z)^{s-k-1}\Laplace^{-1}\frac{(Z_{\mathbb{O'}_{s}}(\lambda) J_{\mathbb{O}^{'E}_{s}}-Z_{\tilde{\mathbb{O}'}_{s}}(\lambda)J_{\tilde{\mathbb{O}}^{'E}_{s}})}{N\,\lambda^{2+s}}(-\overrightarrow{\partial}_z)^{k+1} \right)\nonumber\\
	&+\frac{1}{2}\log\Det\Bigg[I
	-2\left(I+\sum_{k=0}^{s-2}{s\choose k}{s\choose k+2}(-\overrightarrow{\partial}_z)^{s-k-1}\Laplace^{-1}\frac{(Z_{\mathbb{O'}_{s}}(\lambda) J_{\mathbb{O}^{'E}_{s}}-Z_{\tilde{\mathbb{O}'}_{s}}(\lambda)J_{\tilde{\mathbb{O}}^{'E}_{s}})}{N\,\lambda^{2+s}}(-\overrightarrow{\partial}_z)^{k+1} \right)^{-1}\nonumber\\
	& \sum_{k_1=0}^{s_1-2}{s_1\choose k_1}{s_1\choose k_1+2}(-\overrightarrow{\partial}_z)^{s_1-k_1-1}\Laplace^{-1}\frac{Z_{\bar{\mathbb{S}}^{'}_{s_1}}(\lambda)}{N\,\lambda^{2+s_1}}J_{\bar{\mathbb{S}}^{'E}_{s_1}}(-\overrightarrow{\partial}_z)^{k_1+1} \nonumber\\
	&\left(I+\sum_{k_2=0}^{s_2-2}{s_2\choose k_2}{s_2\choose k_2+2}(-\overrightarrow{\partial}_z)^{s_2-k_2-1}\Laplace^{-1}\frac{(Z_{\mathbb{O'}_{s_2}}(\lambda) J_{\mathbb{O}^{'E}_{s_2}}+Z_{\tilde{\mathbb{O}'}_{s_2}}(\lambda)J_{\tilde{\mathbb{O}}^{'E}_{s_2}})}{N\,\lambda^{2+s_2}}(-\overrightarrow{\partial}_z)^{k_2+1} \right)^{-1}\nonumber\\
	& \sum_{k_3=0}^{s_3-2}{s_3\choose k_3}{s_3\choose k_3+2}(-\overrightarrow{\partial}_z)^{s_3-k_3-1}\Laplace^{-1}\frac{Z_{\mathbb{S}'_{s_3}}(\lambda)}{N\,\lambda^{2+s_3}}J_{\mathbb{S}^{'E}_{s_3}}(-\overrightarrow{\partial}_z)^{k_3+1}\Bigg]
\end{align}
that follows from eqs. \eqref{wexplicit}, \eqref{propagator}, \eqref{Wick} and \eqref{Laplacian}.
By the above computation we verify the aforementioned prediction \cite{Bochicchio:2016toi} that the generating functional $\mathcal{W}^E_{asym \, torus}$ of the LO nonplanar Euclidean RG-improved collinear twist-$2$ correlators should inherit the very same loop structure -- specifically, the one of the logarithm of a functional determinant -- of the corresponding nonperturbative object that involves the sum of glueball one-loop diagrams.\par
Hence, the aforementioned asymptotic results strongly constrain the nonperturbative solution of the large-$N$ YM theory and may actually provide an essential guide to find it.

  \newpage
\appendix
  
 \section{Determinants of block matrices} \label{A}
 
The matrix in eq. \eqref{matrixM} has a block structure with noncommutative entries:
  \begin{align}
	& M=
		\begin{pmatrix}A& B\\ C& D \end{pmatrix}
\end{align}
where $A,B,C,D$ are operators. Correspondingly, the determinant of $M$ reads \cite{block}:
\begin{align}
\Det M = \Det (A)\,\Det\left(D-C A^{-1}B\right)
\end{align}
or \cite{powell}:
\begin{align}
	\Det M = \Det (D)\,\Det\left(A-B D^{-1}C\right)
\end{align}
provided that $A^{-1}$ or $D^{-1}$ exist respectively.
The above formulas may be rewritten as:
\begin{align}
	\label{ourdet}
	\Det M =\Det (A)\,\Det(D)\,\Det\left(1-D^{-1}C A^{-1}B\right)
\end{align}
or:
\begin{align}
	\Det M =\Det (D)\,\Det(A)\,\Det\left(1-A^{-1}B D^{-1}C\right)
\end{align}
provided that both $A^{-1}$ and $D^{-1}$ exist.
In our case:
 \begin{align}
	\label{matrixM1}
	&\mathcal{M}^{ab}=\delta^{ab}\scalebox{0.85}{$\begin{pmatrix}i\square-\frac{1}{2}\sum_sJ_{\mathbb{O}_{s}}\otimes\mathcal{Y}_{s-2}^{\frac{5}{2}}-\frac{1}{2}\sum_sJ_{\tilde{\mathbb{O}}_{s}}\otimes\mathcal{H}_{s-2}^{\frac{5}{2}} & -\frac{1}{\sqrt{2}}\sum_s J_{\mathbb{S}_{s}}\otimes\mathcal{Y}_{s-2}^{\frac{5}{2}}\\ -\frac{1}{\sqrt{2}}\sum_sJ_{\bar{\mathbb{S}}_{s}}\otimes\mathcal{Y}_{s-2}^{\frac{5}{2}}&i\square-\frac{1}{2}\sum_sJ_{\mathbb{O}_{s}}\otimes\mathcal{Y}_{s-2}^{\frac{5}{2}}+\frac{1}{2}\sum_sJ_{\tilde{\mathbb{O}}_{s}}\otimes\mathcal{H}_{s-2}^{\frac{5}{2}}  \end{pmatrix}$}
\end{align}
 and we employ eq. \eqref{ourdet}. Hence, up to a trivial constant normalization, we get:
\begin{align}
	\label{detMat}
	\Det \mathcal{M}=&\Det\left(\mathcal{I}+\frac{1}{2}i\square^{-1}J_{\mathbb{O}_{s}}\otimes\mathcal{Y}_{s-2}^{\frac{5}{2}}+\frac{1}{2}i\square^{-1}J_{\tilde{\mathbb{O}}_{s}}\otimes\mathcal{H}_{s-2}^{\frac{5}{2}} \right)\nonumber\\
	&\Det\Big(\mathcal{I}+\frac{1}{2}i\square^{-1}J_{\mathbb{O}_{s}}\otimes\mathcal{Y}_{s-2}^{\frac{5}{2}}-\frac{1}{2}i\square^{-1}J_{\tilde{\mathbb{O}}_{s}}\otimes\mathcal{H}_{s-2}^{\frac{5}{2}}\Big)\nonumber\\
	&\Det\Bigg[\mathcal{I}-\frac{1}{2}\left(\mathcal{I}+\frac{1}{2}i\square^{-1}J_{\mathbb{O}_{s}}\otimes\mathcal{Y}_{s-2}^{\frac{5}{2}}-\frac{1}{2}i\square^{-1}J_{\tilde{\mathbb{O}}_{s}}\otimes\mathcal{H}_{s-2}^{\frac{5}{2}}\right)^{-1}\nonumber\\
	&\quad\quad i\square^{-1}J_{\bar{\mathbb{S}}_{s_1}} \otimes\mathcal{Y}_{s_1-2}^{\frac{5}{2}}\left(\mathcal{I}+\frac{1}{2}i\square^{-1}J_{\mathbb{O}_{s_2}}\otimes\mathcal{Y}_{s_2-2}^{\frac{5}{2}}+\frac{1}{2}i\square^{-1}J_{\tilde{\mathbb{O}}_{s_2}}\otimes\mathcal{H}_{s_2-2}^{\frac{5}{2}} \right)^{-1}\nonumber\\
	&\quad\quad i\square^{-1}J_{\mathbb{S}_{s_3}} \otimes\mathcal{Y}_{s_3-2}^{\frac{5}{2}} \Bigg]
\end{align}
where the sum over repeated spin indices is understood. 

\section{Conformal properties of the standard basis} \label{B}

The gauge-invariant collinear twist-$2$ operators in the light-cone gauge in the standard basis read \cite{BPS1}:
\begin{align} \label{1000}
\nonumber
&\mathbb{O}_{s} = \Tr \partial_{+} \bar{A}(x)(i\overrightarrow{\partial}_++i\overleftarrow{\partial}_+)^{s-2}C^{\frac{5}{2}}_{s-2}\left(\frac{\overrightarrow{\partial}_+-\overleftarrow{\partial}_+}{\overrightarrow{\partial}_++\overleftarrow{\partial}_+}\right)\partial_{+} {A}(x)\\\nonumber
&\tilde{\mathbb{O}}_{s} = \Tr \partial_{+} \bar{A}(x)(i\overrightarrow{\partial}_++i\overleftarrow{\partial}_+)^{s-2}C^{\frac{5}{2}}_{s-2}\left(\frac{\overrightarrow{\partial}_+-\overleftarrow{\partial}_+}{\overrightarrow{\partial}_++\overleftarrow{\partial}_+}\right)\partial_{+} {A}(x) \\\nonumber
&\mathbb{S}_{s} =\frac{1}{\sqrt{2}}\Tr \partial_{+} \bar{A}(x)(i\overrightarrow{\partial}_++i\overleftarrow{\partial}_+)^{s-2}C^{\frac{5}{2}}_{s-2}\left(\frac{\overrightarrow{\partial}_+-\overleftarrow{\partial}_+}{\overrightarrow{\partial}_++\overleftarrow{\partial}_+}\right)\partial_{+} \bar{A}(x)\\
&\bar{\mathbb{S}}_{s} =\frac{1}{\sqrt{2}}\Tr \partial_{+} A(x)(i\overrightarrow{\partial}_++i\overleftarrow{\partial}_+)^{s-2}C^{\frac{5}{2}}_{s-2}\left(\frac{\overrightarrow{\partial}_+-\overleftarrow{\partial}_+}{\overrightarrow{\partial}_++\overleftarrow{\partial}_+}\right)\partial_{+} A(x)
\end{align}
where $C^{\alpha'}_l(x)$ are the Gegenbauer polynomials that are a special case of the Jacobi polynomials \cite{BPS1}:
\begin{equation} \label{GPol}
	C^{\alpha'}_l(x) = \frac{\Gamma(l+2\alpha')\Gamma(\alpha'+\frac{1}{2})}{\Gamma(2\alpha')\Gamma(l+\alpha'+\frac{1}{2})}P_l^{(\alpha'-\frac{1}{2},\alpha'-\frac{1}{2})}(x)
\end{equation}
with the symmetry properties:
\begin{align}
	C_{l}^{\alpha'}(-x)=(-1)^{l}C_{l}^{\alpha'}(x)
\end{align}
They are the restriction, up to perhaps normalization and linear combinations \cite{BPS1}, to the component with maximal-spin projection $s$ along the $p_+$ direction of the balanced,
$\mathbb{O}^{\mathcal{T}=2}_{s}, \tilde{\mathbb{O}}^{\mathcal{T}=2}_{s}$, and unbalanced, $\mathbb{S}^{\mathcal{T}=2}_{s}$, twist-$2$ operators that to the leading order of perturbation theory transform as primary operators with respect to the conformal group \cite{Beisert:2004fv}:
\begin{align} \label{1}
&\mathbb{O}^{\mathcal{T}=2}_{s} \quad=\quad \Tr\, F^\mu_{(\rho_1}\overleftrightarrow{D}_{\rho_2}\ldots \overleftrightarrow{D}_{\rho_{s-1}}F_{\rho_s)\mu}-\,\text{traces}\qquad\qquad \nonumber\\
&\tilde{\mathbb{O}}^{\mathcal{T}=2}_{s} \quad=\quad \Tr\, \tilde{F}^\mu_{(\rho_1}\overleftrightarrow{D}_{\rho_2}\ldots \overleftrightarrow{D}_{\rho_{s-1}}F_{\rho_s)\mu}-\,\text{traces}\qquad\qquad \nonumber\\
&\mathbb{S}^{\mathcal{T}=2}_{s} \quad\,\,=\quad \Tr\, (F_{\mu(\nu}+i\tilde{F}_{\mu(\nu})\overleftrightarrow{D}_{\rho_1}\ldots \overleftrightarrow{D}_{\rho_{s-2}}(F_{\lambda)\sigma}+i\tilde{F}_{\lambda)\sigma})-\,\text{traces}\qquad\qquad 
\end{align}
where the parentheses stand for symmetrization of all the indices in between and the subtraction of the traces ensures that the contraction of any two indices is zero.\par
The above statement about the conformal properties needs some refinements.
In a conformal field theory the propagator of a vector field $V_\mu(x)$ with conformal dimension $1$ is purely longitudinal \cite{makeenko}: 
\begin{equation}
	\langle V_\mu(x)V_\nu(0)\rangle = \mathcal{C}_2 \left(g_{\mu\nu}-2\frac{x_\mu x_\nu}{\rvert x\rvert^2}\right)\frac{1}{\rvert x\rvert^2}=\frac{\mathcal{C}_2}{2}\partial_\mu\partial_\nu\log \rvert x \rvert ^2 
\end{equation}
Therefore, there is no Lorentz-covariant gauge where the field $A_{\mu}$ may be primary with respect to the conformal group to the lowest order of perturbation theory. Yet, $F_{\mu \nu}$ decomposes into the $(1,0)\oplus(0,1)$ representations of the Lorentz group \cite{BPS1}:
\begin{equation}
	\label{lorentza}
	F_{a\dot{a}b\dot{b}} = 2(f_{ab}\epsilon_{\dot{a}\dot{b}} - \epsilon_{ab} f_{\dot{a}\dot{b}})
\end{equation}
where \cite{BPS1}:
\begin{equation}
	\label{adota}
	F_{a\dot{a}b\dot{b}} = \sigma_{a\dot{a}}^\mu\sigma_{b\dot{b}}^\nu F_{\mu\nu}
\end{equation}
and:
\begin{align} 
&f_{ab} = \frac{i}{2}(\sigma^{\mu\nu})_{ab}F_{\mu\nu}\nonumber\\
&f_{\dot{a}\dot{b}} = -\frac{i}{2}(\bar{\sigma}^{\mu\nu})_{\dot{a}\dot{b}}F_{\mu\nu}
\end{align}
with:
\begin{equation}
	\label{dagger}
	\bar{f}_{ab} = f_{\dot{a}\dot{b}}
\end{equation}
It turns out that to the leading order of perturbation theory $f_{ab}$ and $f_{\dot{a}\dot{b}}$ are local primary fields with respect to the conformal group \cite{Beisert:2004fv}.
Consequently, it follows the construction \cite{Beisert:2004fv} in eq. \eqref{1} of the local gauge-invariant operators that are conformal primaries to the leading order, with the caveat above that the corresponding representation of the conformal group does not extend outside the local gauge-invariant sector by including the field $A_{\mu}$ in a covariant gauge.\par
Besides, by projecting on the maximal-spin component along the $p_+$ direction the aforementioned gauge-invariant operators restrict to \cite{BPS1}:
\begin{align} \label{OO}
\nonumber
&\mathbb{O}_{s} = \Tr f_{11}(x)(i\overrightarrow{D}_++i\overleftarrow{D}_+)^{s-2}C^{\frac{5}{2}}_{s-2}\left(\frac{\overrightarrow{D}_+-\overleftarrow{D}_+}{\overrightarrow{D}_++\overleftarrow{D}_+}\right) f_{\dot{1}\dot{1}}(x) \qquad s = 2,4,6,\ldots  \\\nonumber
&\tilde{\mathbb{O}}_{s} = \Tr f_{11}(x)(i\overrightarrow{D}_++i\overleftarrow{D}_+)^{s-2}C^{\frac{5}{2}}_{s-2}\left(\frac{\overrightarrow{D}_+-\overleftarrow{D}_+}{\overrightarrow{D}_++\overleftarrow{D}_+}\right) f_{\dot{1}\dot{1}}(x) \qquad s = 3,5,7,\ldots  \\\nonumber
&\mathbb{S}_{s} =\frac{1}{\sqrt{2}}\Tr f_{11}(x)(i\overrightarrow{D}_++i\overleftarrow{D}_+)^{s-2}C^{\frac{5}{2}}_{s-2}\left(\frac{\overrightarrow{D}_+-\overleftarrow{D}_+}{\overrightarrow{D}_++\overleftarrow{D}_+}\right)f_{11}(x) \qquad s = 2,4,6,\ldots\\
&\bar{\mathbb{S}}_{s} =\frac{1}{\sqrt{2}}\Tr f_{\dot{1}\dot{1}}(x)(i\overrightarrow{D}_++i\overleftarrow{D}_+)^{s-2}C^{\frac{5}{2}}_{s-2}\left(\frac{\overrightarrow{D}_+-\overleftarrow{D}_+}{\overrightarrow{D}_++\overleftarrow{D}_+}\right) f_{\dot{1}\dot{1}}(x)\qquad s = 2,4,6,\ldots
\end{align}
that are primaries with respect to the collinear conformal subgroup $SL(2,R)$ \cite{Belitsky:1998gc}
and reduce in the light-cone gauge to the operators in eq. \eqref{1000} with $f_{11}=- \partial_+ \bar A$. By allowing operator mixing with derivatives of operators with lower spin they may be extended to primary conformal operators to the next-to-leading order in the conformal renormalization scheme \cite{Braun:2003rp} that differs from the $\overline{MS}$ scheme by a finite renormalization.\par
In fact, the fields $f\equiv f_{11}$ and $\bar f \equiv f_{\dot{1}\dot{1}}$ may be realized as primary operators for the collinear conformal subgroup to the leading order of perturbation theory in a noncanonical path-integral quantization of the YM theory in the light-cone gauge by a suitable change of variables \cite{Belitsky:2004sc}:
  \begin{align}
S_{YM}(f,\bar f) = &\int \bar{f}^a \partial_+^{-2}\square f^a+2\frac{g}{\sqrt{N}} f^{abc}(\partial_+^{-1}\bar f^a f^b\bar{\partial}\partial_+^{-2}\bar f^c+\partial_+^{-1}f^a \bar f^b\partial\partial_+^{-2}f^c)\nonumber\\
&-2\frac{g^2}{N}f^{abc}f^{ade}\partial_+^{-1}(\partial_+^{-1}\bar f^b  f^c)\partial_+^{-1}(\partial_+^{-1}f^d \bar{f}^e) \,d^4x
\end{align} 
where:
\begin{equation}
  \langle\mathcal{O}_1(x_1)\ldots \mathcal{O}_n(x_n)\rangle =\frac{1}{Z}\int \mathcal{D}f\mathcal{D}\bar{f}\, e^{iS_{YM}(f,\bar f)} \mathcal{O}_1(x_1)\ldots \mathcal{O}_n(x_n)
 \end{equation}

 \bibliographystyle{JHEP}

\bibliography{mybib} 

\providecommand{\href}[2]{#2}\begingroup\raggedright\begin{thebibliography}{10}

\bibitem{tHooft:1973alw}
G.~'t~Hooft, \emph{{A Planar Diagram Theory for Strong Interactions}},
  \href{http://dx.doi.org/10.1016/0550-3213(74)90154-0}{\emph{Nucl. Phys. B}
  {\bf 72} (1974) 461}.

\bibitem{Witten:1979kh}
E.~Witten, \emph{{Baryons in the 1/n Expansion}},
  \href{http://dx.doi.org/10.1016/0550-3213(79)90232-3}{\emph{Nucl. Phys. B}
  {\bf 160} (1979) 57--115}.

\bibitem{Migdal:1977nu}
A.~A. Migdal, \emph{{Multicolor QCD as Dual Resonance Theory}},
  \href{http://dx.doi.org/10.1016/0003-4916(77)90181-6}{\emph{Annals Phys.}
  {\bf 109} (1977) 365}.

\bibitem{Bochicchio:2016toi}
M.~Bochicchio, \emph{{An asymptotic solution of Large-N QCD, for the glueball
  and meson spectrum and the collinear S-matrix}},
  \href{http://dx.doi.org/10.1063/1.4949387}{\emph{AIP Conf. Proc.} {\bf 1735}
  (2016) 030004}.

\bibitem{MB1}
M.~Bochicchio, \emph{{On the geometry of operator mixing in massless QCD-like
  theories}},
  \href{http://dx.doi.org/10.1140/epjc/s10052-021-09543-5}{\emph{Eur. Phys. J.
  C} {\bf 81} (2021) 749}, [\href{https://arxiv.org/abs/2103.15527}{{\tt
  2103.15527}}].

\bibitem{BPS1}
M.~Bochicchio, M.~Papinutto and F.~Scardino, \emph{{n-point correlators of
  twist-2 operators in SU(N) Yang-Mills theory to the lowest perturbative
  order}}, \href{http://dx.doi.org/10.1007/JHEP08(2021)142}{\emph{JHEP} {\bf
  08} (2021) 142}, [\href{https://arxiv.org/abs/2104.13163}{{\tt 2104.13163}}].

\bibitem{Beisert:2004fv}
N.~Beisert, G.~Ferretti, R.~Heise and K.~Zarembo, \emph{{One-loop QCD spin
  chain and its spectrum}},
  \href{http://dx.doi.org/10.1016/j.nuclphysb.2005.04.004}{\emph{Nucl. Phys. B}
  {\bf 717} (2005) 137--189}, [\href{https://arxiv.org/abs/hep-th/0412029}{{\tt
  hep-th/0412029}}].

\bibitem{Robertson:1990bf}
D.~G. Robertson and F.~Wilczek, \emph{{Anomalous dimensions of anisotropic
  gauge theory operators}},
  \href{http://dx.doi.org/10.1016/0370-2693(90)90731-K}{\emph{Phys. Lett. B}
  {\bf 251} (1990) 434--438}.

\bibitem{BPSN}
To appear in arXiv.

\bibitem{Belitsky:1998gc}
A.~V. Belitsky and D.~Mueller, \emph{{Broken conformal invariance and spectrum
  of anomalous dimensions in QCD}},
  \href{http://dx.doi.org/10.1016/S0550-3213(98)00677-4}{\emph{Nucl. Phys. B}
  {\bf 537} (1999) 397--442}, [\href{https://arxiv.org/abs/hep-ph/9804379}{{\tt
  hep-ph/9804379}}].

\bibitem{Braun2}
V.~M. Braun, A.~N. Manashov, S.~Moch and M.~Strohmaier, \emph{{Three-loop
  evolution equation for flavor-nonsinglet operators in off-forward
  kinematics}}, \href{http://dx.doi.org/10.1007/JHEP06(2017)037}{\emph{JHEP}
  {\bf 06} (2017) 037}, [\href{https://arxiv.org/abs/1703.09532}{{\tt
  1703.09532}}].

\bibitem{Braun3}
V.~M. Braun, Y.~Ji and A.~N. Manashov, \emph{{Two-loop evolution equation for
  the B-meson distribution amplitude}},
  \href{http://dx.doi.org/10.1103/PhysRevD.100.014023}{\emph{Phys. Rev. D} {\bf
  100} (2019) 014023}, [\href{https://arxiv.org/abs/1905.04498}{{\tt
  1905.04498}}].

\bibitem{Aglietti:2021bem}
U.~Aglietti, M.~Becchetti, M.~Bochicchio, M.~Papinutto and F.~Scardino,
  \emph{{Operator mixing, UV asymptotics of nonplanar/planar $2$-point
  correlators, and nonperturbative large-$N$ expansion of QCD-like theories}},
  \href{https://arxiv.org/abs/2105.11262}{{\tt 2105.11262}}.

\bibitem{Belitsky:2004sc}
A.~V. Belitsky, S.~E. Derkachov, G.~P. Korchemsky and A.~N. Manashov,
  \emph{{Dilatation operator in (super-)Yang-Mills theories on the
  light-cone}},
  \href{http://dx.doi.org/10.1016/j.nuclphysb.2004.11.034}{\emph{Nucl. Phys. B}
  {\bf 708} (2005) 115--193}, [\href{https://arxiv.org/abs/hep-th/0409120}{{\tt
  hep-th/0409120}}].

\bibitem{BB}
M.~Becchetti and M.~Bochicchio, \emph{{Operator mixing in massless QCD-like
  theories and Poincar\`e\textendash{}Dulac theorem}},
  \href{http://dx.doi.org/10.1140/epjc/s10052-022-10551-2}{\emph{Eur. Phys. J.
  C} {\bf 82} (2022) 866}, [\href{https://arxiv.org/abs/2103.16220}{{\tt
  2103.16220}}].

\bibitem{Braun:2003rp}
V.~M. Braun, G.~P. Korchemsky and D.~Mueller, \emph{{The Uses of conformal
  symmetry in QCD}},
  \href{http://dx.doi.org/10.1016/S0146-6410(03)90004-4}{\emph{Prog. Part.
  Nucl. Phys.} {\bf 51} (2003) 311--398},
  [\href{https://arxiv.org/abs/hep-ph/0306057}{{\tt hep-ph/0306057}}].

\bibitem{block}
S.~R. Garcia and R.~A. Horn, \emph{Block matrices in linear algebra},
  \href{http://dx.doi.org/10.1080/10511970.2019.1567214}{\emph{PRIMUS} {\bf 30}
  (2020) 285--306}.

\bibitem{powell}
P.~D. Powell, \emph{{Calculating Determinants of Block Matrices}},
  \href{https://arxiv.org/abs/1112.4379}{{\tt 1112.4379}}.

\bibitem{makeenko}
Y.~M. Makeenko, \emph{On conformal operators in quantum chromodynamics},  tech.
  rep., USSR, 1980.

\end{thebibliography}\endgroup

\end{document}